\documentclass[12pt]{iopart}
\usepackage{epsfig}
\newcommand{\ii}{\mathrm{i}}
\newcommand{\ee}{\mathrm{e}}
\newcommand{\dd}{\mathrm{d}}
\newcommand{\sn}{\mathrm{sn}}
\newcommand{\dn}{\mathrm{dn}}
\newcommand{\cn}{\mathrm{cn}}
\newcommand{\K}{\mathbf{K}}
\newcommand{\E}{\mathbf{E}}
\newcommand{\EK}{\frac{\E}{\K}}
\newcommand{\re}{\mathrm{Re}}
\newcommand{\w}{\omega}
\begin{document}
\title[Updating the Gross-Neveu phase diagram]{From relativistic quantum fields\\
to condensed matter and back again: \\Updating the Gross-Neveu phase diagram}
\author{Michael Thies}
\address{Institut f\"ur Theoretische Physik III, Universit\"at Erlangen-N\"urnberg, Staudtstr.~7,
91058 Erlangen, Germany}
\ead{thies@theorie3.physik.uni-erlangen.de}
\begin{abstract}
During the last few years, the phase diagram of the large $N$ Gross-Neveu model in 1+1 dimensions
at finite temperature and chemical potential has undergone a major revision. Here we present a
streamlined account of this development, collecting the most important results.
Quasi-one-dimensional condensed matter systems like conducting polymers provide real physical 
systems which can be approximately described by the Gross-Neveu model and have played some role
in establishing its phase structure.  
The kink-antikink phase found at low temperatures is closely related to inhomogeneous 
superconductors in the Larkin-Ovchinnikov-Fulde-Ferrell phase.
With the complete phase diagram at hand, the Gross-Neveu model can now serve as a firm testing ground for 
new algorithms and theoretical ideas.

\end{abstract}
\maketitle
\section{Introduction}

In its original form, the Gross-Neveu (GN) model \cite{L1} is a relativistic, renormalizable quantum field theory (QFT) of $N$
species
of self-interacting fermions in 1+1 dimensions with Lagrangian
\begin{equation}
{\cal L} = \sum_{n=1}^N \bar{\psi}^{(n)} ({\rm i}\gamma^{\mu}\partial_{\mu} - m_0)\psi^{(n)} + \frac{1}{2} g^2 \left(\sum_{n=1}^N 
\bar{\psi}^{(n)}\psi^{(n)}\right)^2.
\label{1.1}
\end{equation}
The bare mass term $\sim m_0$ explicitly breaks the discrete chiral symmetry $\psi\to \gamma^5 \psi$ of the massless model.
The interaction term can be generalized to acquire a continuous chiral symmetry as in the Nambu--Jona-Lasinio (NJL) 
model, but we shall not consider this option here.
As far as the phase diagram is concerned, the 't~Hooft limit $N\to \infty$, $g^2 \sim 1/N$ is most instructive.
The reason is the following. In 1+1 dimensions, classic no-go theorems forbid 
spontaneous breaking of a continuous symmetry at zero temperature or of a discrete symmetry at finite
temperature \cite{L2,L3}. Although mean-field theory often predicts these effects, fluctuations are expected to 
destroy any long range order. Provided that one defines the model by letting $N \to \infty$ before taking
the thermodynamic limit, these fluctuations are suppressed and a richer phase structure becomes accessible \cite{L4}.
We follow this line of reasoning here. 

Like in a previous pedagogical review article \cite{L5} which the present paper is meant to update, we work canonically, 
couching the theory in the 
language of the relativistic Hartree-Fock (HF) approximation. It is formally equivalent to the semi-classical 
functional integral approach
but conceptually simpler. Physically, it emphasizes the fact that the Dirac sea is an interacting many-fermion system
\cite{L6,L7}, thereby making it somewhat easier to switch back and forth between the relativistic QFT and
condensed matter physics literature.  

Let us now briefly recall the motivation for studying the Lagrangian (\ref{1.1}). QFT models for which one
can construct the exact renormalized phase diagram are extremely scarce and worth investigating on theoretical grounds,
even if they are far from being realistic. As a matter of fact, in spite of its simple Lagrangian, the GN model shares
many non-trivial properties with quantum chromodynamics (QCD), notably asymptotic freedom,
dimensional transmutation, meson and baryon bound states, chiral symmetry breaking in the vacuum as well as its
restoration at high temperature and density. 
Perhaps even more surprising and less widely appreciated is the fact that GN type models have enjoyed
considerable success in describing a variety of
quasi-one-dimensional condensed matter systems such as the Peierls-Fr\"ohlich model,   
conducting polymers like polyacetylene, or inhomogeneous superconductors. 

The title of this paper has the following background: Originally, the kink
and kink-antikink baryons first derived in QFT by Dashen, Hasslacher and Neveu \cite{L8} have been useful 
for understanding the role of solitons and polarons in electrical conductivity properties of doped polymers \cite{L9}. 
In the following years, a lot of progress was made in condensed matter theory towards understanding polaron
crystal structures. This in turn 
helped us to construct the full phase diagram of the relativistic QFT. 
However, in order not to confuse the issues, we shall first discuss the GN model  without
reference to condensed matter physics and comment on the relationship only towards the end.

Our last remark about the motivation concerns the restriction to 1+1 dimensions. At first glance, this looks
very unrealistic. However, most of the work discussed below has to do with finite chemical potential. In the
presence of a Fermi surface, a kind of dimensional reduction takes place which has been exploited for instance
in high density effective theory  (HDET) to QCD (see \cite{L10} for a review). It also manifests itself through the well-known
fact that
Cooper pairing in
superconductivity occurs for arbitrarily weak attraction, just like in one space dimension.
Hence, from the physics point of view, the restriction to 1+1 dimensions is perhaps a better idea in the
presence of fermionic matter than naively thought.

The Lagrangian (\ref{1.1}) has two bare parameters, $g^2$ and $m_0$. In the process
of regularization and renormalization, all observables can be expressed in terms of two physical parameters $m$
 and $\gamma$.
The relation to the bare quantities and the ultra-violet (UV) cutoff $\Lambda$ is given by the vacuum gap equation
\begin{equation}
\frac{\pi}{Ng^2} = \gamma + \ln \frac{\Lambda}{m}, \qquad  \gamma:= \frac{\pi}{Ng^2}\frac{m_0}{m}.
\label{1.2}
\end{equation}
Whereas the physical fermion mass $m$ in the vacuum merely provides the overall mass scale and can be set equal to 1,
the second parameter $\gamma$
(called ``confinement parameter" in condensed matter physics) parametrizes different physical theories. It measures
 the amount
of explicit chiral symmetry breaking and vanishes in the massless ($m_0=0$) limit. 
Notice that it can also be expressed in terms of the physical fermion masses at bare mass $m_0$ and in the chiral limit,
\begin{equation}
\gamma = \ln \left(\frac{m[m_0]}{m[0]}\right).
\label{1.3}
\end{equation}

The main subject of this paper is the full phase diagram of the massive GN model
as a function of temperature $T$, chemical potential $\mu$ and confinement parameter $\gamma$. 
The original works date from 1985 for the massless \cite{L11} and 1995 
for the massive case \cite{L12} (see also \cite{L12a} for earlier, partial results). As has become clear by now, 
the assumption that the scalar condensate $\langle \bar{\psi}\psi\rangle$ is
spatially homogeneous made in these works is too restrictive. It misses important physics closely related to the existence
of kink-antikink baryons in the GN model. In the present paper we describe how this problem has been cured.
This is not a small correction, but requires a substantial new effort. 
A homogeneous condensate acts like a mass and reduces
the thermodynamics essentially to that of a massive, free relativistic Fermi gas, the mass being determined 
self-consistently. By contrast, solving the Dirac equation with a periodic potential and establishing 
self-consistency is a highly non-trivial task, even in 1+1 dimensions. Due to a number of lucky circumstances,
this is nevertheless possible in the case at hand, to a large extent even analytically. 

In our recent papers we had to proceed from special cases to more general ones (treating $m_0=0$ before
$m_0 \neq 0$, $T=0$ before $T\neq 0$) in order to reduce the problem to manageable size.
Now that the original technical difficulties have been overcome, we 
invert this chronological order. We start from the most general case, i.e., the massive GN model at finite
temperature and chemical potential,  and specialize further wherever additional 
analytical results or physics insights are available. In this way we can hopefully convey a more coherent picture
of what has been learned in the meantime.

This paper is organized as follows. In section 2, we remind the reader of the state of the art around the year 2000 and
recall our criticism on the widely accepted phase diagram of the GN model at that time. We motivate the
analytical form of the correct self-consistent HF potential in section 3 and sketch the calculation of the grand
canonical potential and the proof of self-consistency in section 4. Section 5 exhibits the revised phase diagram
``in full glory". In the following four sections we have added various results in the form of figures and analytical formulas which
have been obtained in simpler special cases, letting one  or both of the parameters $T$ and $\gamma$ go to 
zero. If we take the low density limit $\rho \to 0$ at zero temperature, we also recover the properties of 
individual baryons as summarized in section 10. Section 11 contains a brief overview of the many
cross-relations between the GN model as a relativistic QFT and the theory of certain quasi-one-dimensional
condensed matter systems. It is followed by our concluding remarks in section 12.
 
\section{Reminder of the situation five years ago}

Let us briefly recall the phase diagram of the GN model as it had been widely accepted by the year 2000.
This is necessary to put the recent development into perspective. Besides, we will see that the results 
remain valid in some regions of the phase diagram, including some critical curves. The most basic quantity at finite
temperature $T$ and chemical  potential $\mu$ is 
the grand canonical potential. In the large $N$ limit, it can be computed either canonically via
the relativistic HF approach or by using semi-classical functional integral methods. 
In the more popular path integral approach, one introduces an auxiliary bosonic field $\sigma$  and integrates out the
 fermions exactly
(Gaussian integral over Grassmann variables), then applies a saddle point approximation to the remaining bosonic
functional integral.
In the original analysis at $\gamma=0$ \cite{L11} and its generalization to finite bare fermion masses \cite{L12},
it has been assumed that the classical value of $\sigma$ (i.e., the saddle point) is spatially constant, yielding a
($\mu,T$)-dependent dynamical fermion mass $M$. This mass is found by minimizing the renormalized
grand canonical potential density
\begin{equation}
\fl
\Psi = \frac{M^2}{2\pi}\left( \ln M-\frac{1}{2}\right)+\gamma\left(\frac{M^2}{2\pi}-\frac{M}{\pi}\right)
- \frac{1}{\beta \pi} \int_0^{\infty} {\rm d}q \ln \left[ \left( 1+ {\rm e}^{-\beta(E-\mu)}\right) \left( 1+
 {\rm e}^{-\beta(E+\mu)} \right)\right]
\label{2.1}
\end{equation}
($\beta=1/T, m=1, E=\sqrt{q^2+M^2}\,$) with respect to $M$. Equivalently, one can solve 
the self-consistency condition for the (thermal) fermion condensate,
\begin{equation}
M=m_0  -Ng^2 \langle\bar{\psi}\psi\rangle_{\rm th}.
\label{2.2}
\end{equation}
Depending on the parameters ($\gamma,\mu,T$), $\Psi$ possesses
one or two local minima with the possibility of a first order phase transition. 
The phase diagram in the chiral limit $\gamma=0$ is shown in figure 1, whereas the 3d-plot in figure 2
exhibits the phase structure in the full ($\gamma,\mu,T$)-space. 
Let us first look at figure 1.
The line AB is a critical line of second order transitions
(the thermodynamic potential changes from one minimum to a
maximum and a minimum).
The point B is a tricritical point located at
$1/\beta = 0.3183,\ \mu=0.6082$ (all numbers here 
are in units of $m$); it separates the second order line from a
first order line BD
along which the potential has two degenerate minima and a maximum.
The endpoint D lies at $\mu=1/\sqrt{2}$ where the $T=0$ phase
transition occurs.
Lines BC and BE are boundaries
of metastability; when crossed,
the potential acquires or loses a second minimum.
In region OABD,
chiral symmetry is broken and the fermions are massive;
the outside region has unbroken chiral symmetry
and massless fermions.
As the parameter $\gamma$ is switched on,
the 2nd order line AB disappears in favor of a cross-over where the fermion mass changes rapidly, but smoothly.
The first order line on the other hand survives, ending at a critical point. If plotted against $\gamma$, these
critical points lie on the third curve P$_{\rm t}$C emanating from the tricritical point as shown in figure 2. For $\gamma>0$
the effective fermion mass $M$ is everywhere different from zero. If one crosses the shaded critical ``sheet" in figure 2,
the mass changes discontinuously, dropping with increasing chemical potential.

\begin{figure}[ht]
\begin{center}
\epsfig{file=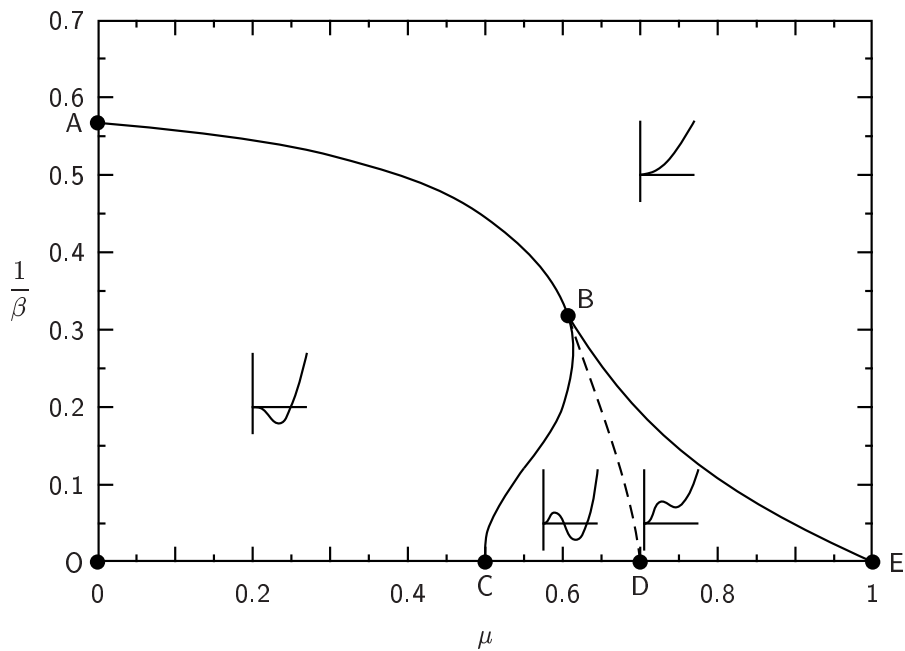,width=12cm}
\caption{``Old" phase diagram of the massless GN model, assuming
unbroken translational symmetry. For a discussion, see main text.
Units of $m$; adapted from reference \cite{L11}.}
\end{center}
\end{figure}

\begin{figure}[ht]
\begin{center}
\epsfig{file=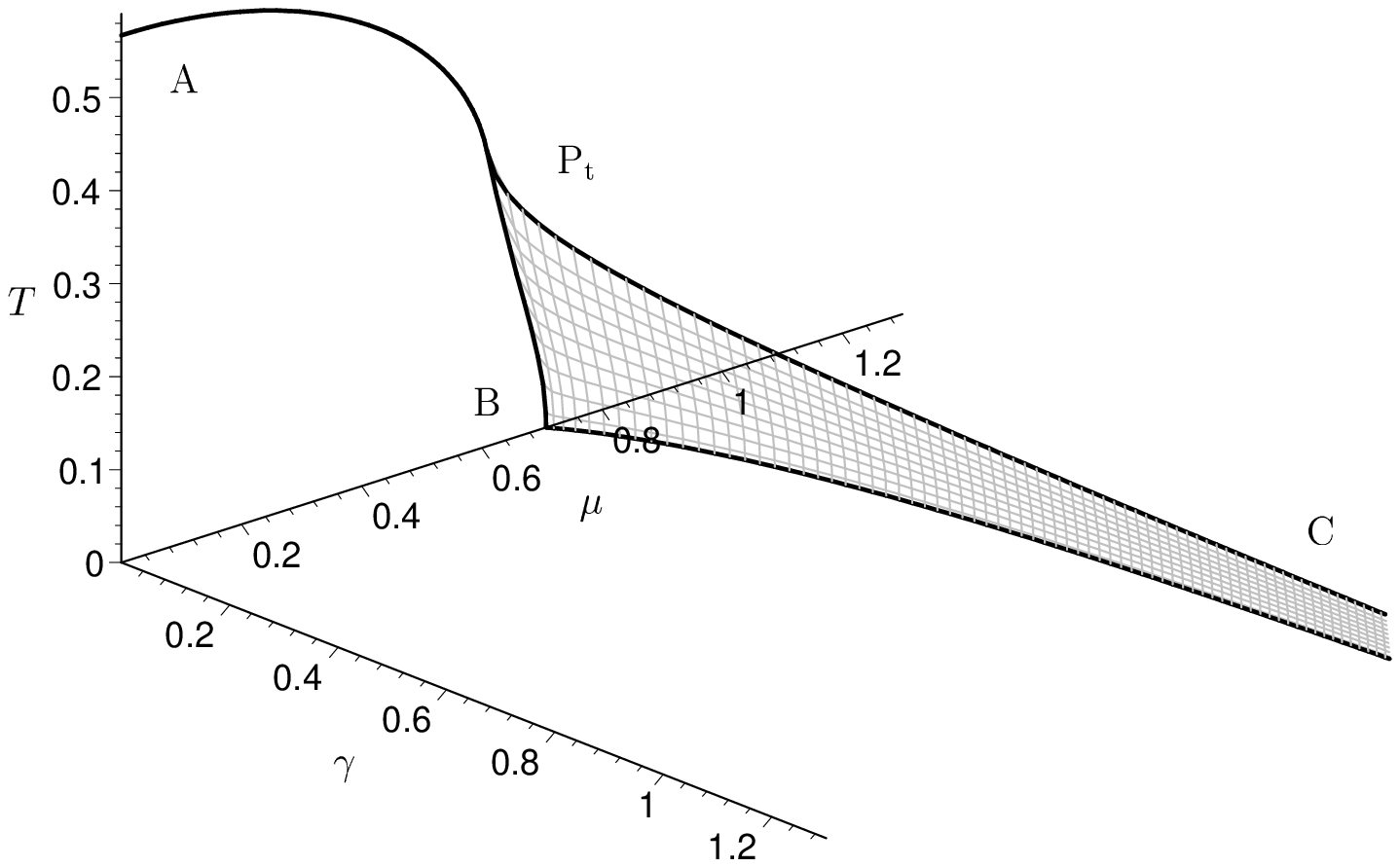,width=8cm,angle=270}
\vskip 1.0cm
\caption{``Old" phase diagram of the massive GN model as a function of $\gamma,\mu,T$, assuming 
unbroken translational symmetry \cite{L12,L13a,L13b}. The various phase boundaries are explained in the main text.
The $\gamma=0$ plane corresponds to figure 1, the tricritical point P$_{\rm t}$ to the point B there.}
\end{center}
\end{figure}

These results looked rather convincing and indeed have been confirmed repeatedly by other authors, see e.g. \cite{L14,L15}.
Nevertheless, they suffer from one disease which was pointed out in reference \cite{L5}.
The GN model possesses baryons (multi-fermion bound states) due to binding effects which are not $1/N$ suppressed.
Consider the low density limit of baryonic matter at $T=0$. Since the baryon-baryon interaction is known 
to be short-ranged and repulsive, one expects widely spaced baryons in the low density limit, implying the 
following slope of the energy density at the origin,
\begin{equation}
\left. \frac{\partial {\cal E}}{\partial \rho}\right|_{\rho=0} = M_B.
\label{2.3}
\end{equation}
Here, $M_B$ is the baryon mass.
Conversely, since equation (\ref{2.3}) is just the chemical potential at $T=0$, the phase
transition at $T=0$ should occur at the critical chemical potential $\mu=M_B$. The phase diagrams in figures 1 and 2
are in conflict with this expectation.
At $\gamma=0$ for instance, the first order phase transition 
takes place at $\mu=1/\sqrt{2}$, whereas the baryon mass is $M_B=2/\pi$ (always taking out a trivial factor of $N$). 
At the time of writing the article \cite{L5}, it was already clear that the problem had to do with the assumption
of translational invariance. Indeed,
if the single baryon breaks translational invariance, there is no good reason why this symmetry should not
be broken at finite baryon densities as well. (One can actually understand the value of $1/\sqrt{2}$ of the 
critical chemical potential in terms
of a droplet model for baryons, characteristic for the mixed phase at low density. The argument which
was first developed in the context of the GN model with continuous chiral symmetry \cite{L16} applies here as well.)

Incidentally, a related observation had already been made in 1987 by Karsch {\em et al.} whose numerical
lattice calculation gave first hints that kink-antikink configurations play a role in the phase diagram at low density \cite{L17}.
However the authors concluded erroneously that there was a problem with the mean field approach, an opinion
which has often been voiced in the literature since then. This calculation was too crude to determine the 
phase diagram quantitatively, in particular the statements about the order of the phase transitions are no longer tenable,
as we now know.
 
Unfortunately, it is not easy to relax the assumption of a constant potential, since this requires a major 
new effort. This can already be judged from the rather involved single baryon problem \cite{L8} which must be contained 
as a limit in the full calculation. Which potential $S(x)$ should replace the constant mass $M$ so as to cure the aforementioned
discrepancy? This will be answered in the next section.

\section{How to choose the right ansatz for the scalar potential}

Our strategy within the relativistic HF approach is extremely simple in principle: guess the functional form of the
HF potential and verify its self-consistency.
At finite density we expect the appearance of a regular array of baryons, hence
we will be searching for periodic solutions. But what guidance do we have
to pick the right ansatz? Is there any reason at all to assume that the exact potential can be given in closed analytical form?  
The following heuristic considerations should be of some help.

Owing to the $(\bar{\psi}\psi)^2$-interaction in the Lagrangian (\ref{1.1}), the Dirac-HF equation for the 
GN model assumes the form
\begin{equation}
\label{3.1}
\left( -{\rm i} \gamma^5   \frac{\partial}{\partial x} + \gamma^0 S(x) \right)\psi(x)=E \psi(x) 
\end{equation}
with real scalar potential $S(x)$. 
The representation 
\begin{equation}
\gamma^0 = - \sigma_1 , \quad \gamma^1={\rm i}\sigma_3, \quad \gamma^5=\gamma^0\gamma^1 = - \sigma_2 
\label{3.2}
\end{equation}
of the $\gamma$-matrices 
has the advantage that the equations for the upper and lower components $\phi_\pm$ of the Dirac spinor $\psi$ can  
be decoupled by simply applying the Dirac Hamiltonian twice,
\begin{equation}
\label{3.3}
\left( -  \frac{\partial^2}{\partial x^2} \mp  \frac{\partial}{\partial x} S+ S^2\right) \phi_{\pm} = E^2 \phi_{\pm} .
\end{equation}
Equation (\ref{3.3}) states that the Schr\"odinger-type Hamiltonians with potentials
$U_{\pm}= S^2 \pm S'$ have the same spectra, a textbook example of supersymmetric quantum mechanics with
 superpotential $S$.
$S(x)$ in turn depends on the eigenfunctions $\psi_{\alpha}$ and eigenvalues $E_{\alpha}$ through a self-consistency
relation, 
\begin{equation}
- \frac{1}{Ng^2} (S(x)-m_0) = \langle \bar{\psi}\psi \rangle_{\rm th} = 
\sum_{\alpha} \bar{\psi}_{\alpha}(x)\psi_{\alpha}(x)\frac{1}{{\rm e}^{\beta(E_{\alpha}-\mu)}+1}.
\label{3.4}
\end{equation}
It generalizes equation (\ref{2.2}) to an $x$-dependent scalar potential.

Let us first recall the results for single baryons in the massive GN model \cite{L18,L19}. The scalar potential for a baryon
 has the form
\begin{equation}
S(x) =  1+y \left[ \tanh(yx-c_0)-\tanh (yx+c_0)\right],
\qquad
c_0 = \frac{1}{2} \,{\rm artanh}\, y.
\label{3.5}
\end{equation}
The parameter $y$ depends on the bare fermion mass and the number of valence fermions. Here
the corresponding isospectral potentials $U_{\pm}$ in the 2nd order equation (\ref{3.3}) are given
by the simplest P\"oschl-Teller potential \cite{L20}
\begin{equation}
S^2 \pm S' = - \frac{2 y^2}{\cosh^2 (yx \pm c_0)}
\label{3.6}
\end{equation}
and differ only by a translation in space.
The distinguishing feature of potential (\ref{3.6}) is the fact that it is reflectionless; indeed, this is the unique reflectionless
potential with a single bound state. It is well known that static solutions of the GN model must correspond to
reflectionless Schr\"odinger potentials \cite{L8,L21}. The fact that the ansatz (\ref{3.5}) leads to self-consistency
is therefore quite plausible.  

Take now a lattice of infinitely many, equidistant P\"oschl-Teller potential wells. As pointed out in references \cite{L22,L23}, 
the lattice sum can be performed yielding a Lam\'e-type potential,
\begin{equation}
\sum_{n=-\infty}^{\infty} \frac{1}{\cosh^2(x-nd)} = \left( \frac{2\kappa {\bf K'}}{\pi}\right)^2 \left\{ \frac{\bf E'}{\kappa^2{\bf K'}}
- {\rm sn}^2\left( \frac{2{\bf K'}}{\pi} x \right) \right\}.
\label{3.7}
\end{equation}
(${\bf K},{\bf E}$ denote complete elliptic integrals of first and second kind,
${\bf K'},{\bf E'}$ the complementary ones with argument $\kappa'=\sqrt{1-\kappa^2}$, {\rm sn} a Jacobi elliptic function of 
modulus $\kappa$). Comparing the 
spatial period of both sides of equation (\ref{3.7}), we can relate $d$ and $\kappa$ via
\begin{equation}
d = \pi \frac{\bf K}{\bf K'}.
\label{3.8}
\end{equation}
How does the fact that the single potential wells (\ref{3.6}) are reflectionless manifest itself in the periodic
extension (\ref{3.7})?
This has been answered long time ago \cite{L24,L25}: 
The periodic potential has a single gap (or, in general, a finite number of gaps), in contrast to 
generic periodic potentials with infinitely many gaps. Thus reflectionless potentials generalize to
``finite band potentials" as one proceeds from a single well to a periodic array.
In the same way as the sech$^2$-potential is the unique reflectionless potential with one bound state,
the ${\rm sn}^2$-potential is the unique single band potential.
Guided by these considerations, let us try to find the most general superpotential of the  (single gap) Lam\'e potential.
After a scale transformation
\begin{equation}
S(x) = A\tilde{S}(\xi) \qquad \xi =Ax  ,
\label{3.9}
\end{equation}
$U_{\pm}(x)=A^2 \tilde{U}_{\pm}(\xi)$ should assume the form of the Lam\'e potential plus constant, up to a possible shift
of the rescaled coordinate ($\xi_+=\xi+b$),
\begin{eqnarray}
\tilde{U}_+ \ = \ \tilde{S}^2 + \tilde{S}' &=& 2 \kappa^2 {\rm sn}^2\xi_+ + \eta,
\nonumber \\
\tilde{U}_- \ = \ \tilde{S}^2 - \tilde{S}' &=& 2 \kappa^2 {\rm sn}^2 \xi + \eta,
\label{3.10}
\end{eqnarray}
or, equivalently,
\begin{eqnarray}
\tilde{S}^2 &=&  \kappa^2 \left( {\rm sn}^2 \xi_+ + {\rm sn}^2 \xi \right) + \eta,
\nonumber \\
\tilde{S}' & = & \kappa^2 \left({\rm sn}^2\xi_+ - {\rm sn}^2 \xi \right).
\label{3.11}
\end{eqnarray}
Let us try to solve equations (\ref{3.11}) for $\tilde{S}$ and $\eta$. 
We differentiate the upper equation (\ref{3.11}) using
\begin{equation}
\left({\rm sn} \, \xi \right)' =  {\rm cn}\, \xi \, {\rm dn}\, \xi
\label{3.12}
\end{equation}
and divide the result by the lower equation, obtaining 
\begin{equation}
\label{3.13}
\tilde{S}(\xi) = \frac{\sn\,\xi_{+} \cn\, \xi_{+} \dn\, \xi_{+} + \sn\,\xi \,\cn\,\xi \,\dn\,\xi}{\sn^2\xi_{+}-\sn^2\xi}  .
\end{equation}
By specializing equation (\ref{3.11}) to $\xi=0$ we can also determine the constant $\eta$, 
\begin{equation}
\eta = \frac{1}{{\rm sn}^2 b}-1-\kappa^2. 
\label{3.15}
\end{equation}
Incidentally, $\tilde{S}$ can be cast into the somewhat simpler form
\begin{equation}
\tilde{S}(\xi) = \kappa^2 \sn\, b \, \sn\, \xi \, \sn(\xi+b) + \frac{\cn \,b \, \dn\, b}{\sn\, b}  .
\label{3.14}
\end{equation}
The scale factor $A$ in equation (\ref{3.9}) is not constrained  
by these considerations, so that the final answer for $S(x)$ depends on three real parameters $A,\kappa,b$.
Thus we conclude that the potential (\ref{3.9},\ref{3.13}) is the most general Dirac potential
leading to a single gap Lam\'e potential (plus constant) in the corresponding 2nd order equations.
This explains why it is a good starting point for finding periodic, static solutions,
and has turned out to be the key to the phase diagram of the massive GN model.
Incidentally, we did not derive the ansatz for $S(x)$ in this way originally, but took it over from
a mathematically closely related problem which had already been solved in condensed matter physics, the bipolaron crystal
in non-degenerate conducting polymers. We will have more to say about this relationship in section 11.
Notice also that the simple and intuitive relation between the single baryon and the crystal exhibited in
equation (\ref{3.7}) is deeply hidden in the corresponding Dirac potentials (\ref{3.5}) and (\ref{3.9},\ref{3.13})
due to the non-linear, non-local relationship between $S$ and $U_{\pm}$.

Inserting our ansatz into equation (\ref{3.3}), we arrive by construction at the single gap Lam\'e equation in the form
\begin{equation}
 \left(-\frac{\partial^2}{\partial \xi^2} + 2\kappa^2 \sn^2(\xi+(b\mp b)/2)\right)\phi_{\pm}={\cal E} \phi_{\pm} .
\label{3.16}
\end{equation}
Using equation (\ref{3.15}), the relation between Dirac eigenvalues $E$ and Lam\'e eigenvalues ${\cal E}$ is
\begin{equation}
\mathcal{E} = \frac{E^2}{A^2} - \frac{1}{{\rm sn}^2 b} +1+ \kappa^2  .
\label{3.17}
\end{equation}
The solutions of  equation (\ref{3.16}) are well known since reference \cite{L26}. 
The eigenfunctions are fairly complicated, although they can still be given in closed form,
\begin{equation}
\phi_+(\xi)=\mathcal{N} \frac{\mathrm{H}(\xi+\alpha)}{\Theta(\xi)}\mathrm{e}^{- \mathrm{Z}(\alpha) \xi}.
\label{3.18}
\end{equation}
Here, $\mathrm{H}$, $\Theta$ and $\mathrm{Z}$ are the Jacobi eta, theta and zeta function, 
respectively. The parameter $\alpha$ is related to the energy and Bloch momentum. An example of the
dispersion relation is shown in figure 3. 

\begin{figure}[ht]
\begin{center}
\epsfig{file=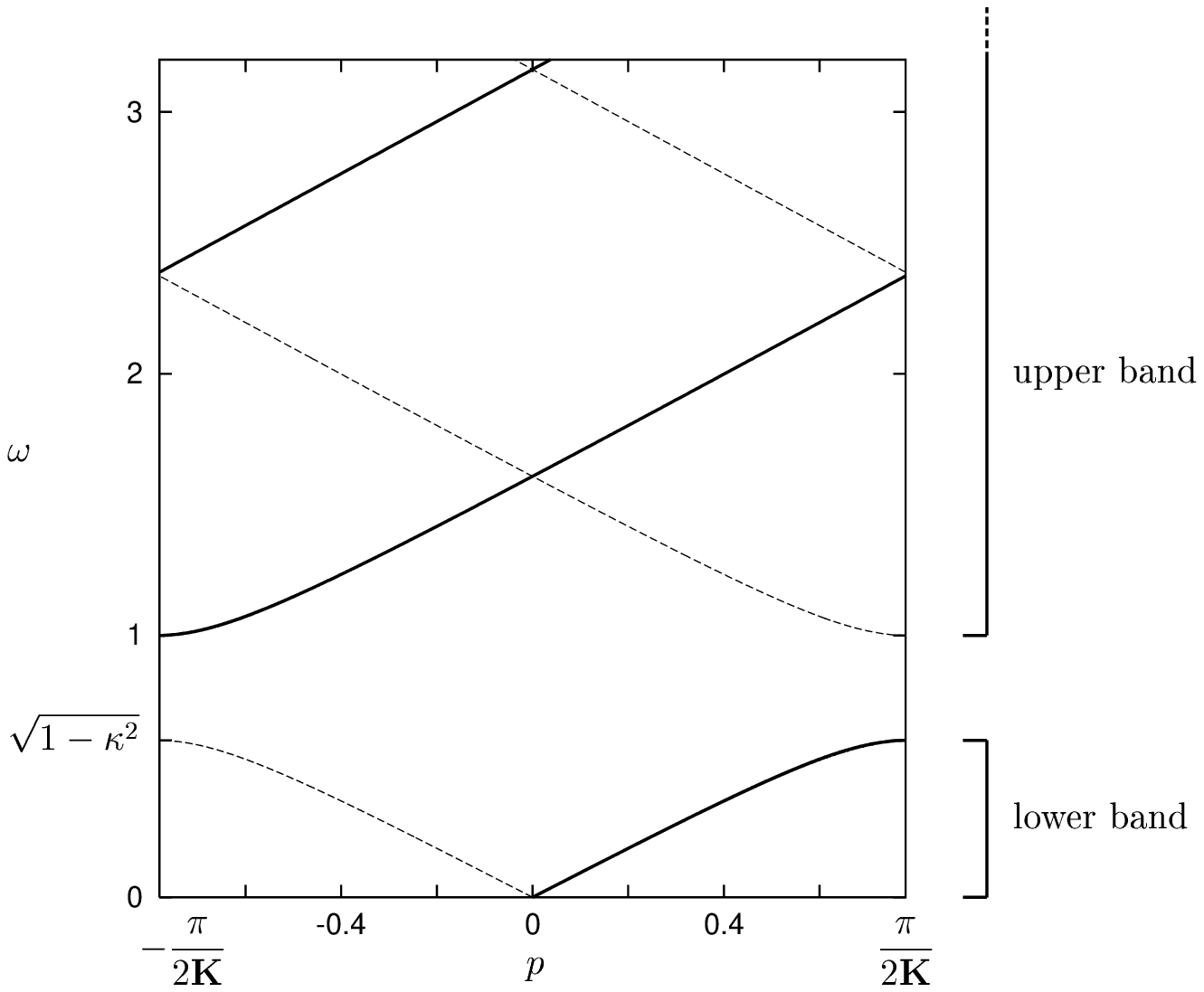,width=10cm}
\caption{The dispersion relation of the Lam\'e potential for $\kappa=0.8$, showing a single gap;
$\omega=E/A$ and $p=q/A$ are (reduced) energy and Bloch momentum, respectively (from reference \cite{L27}).}
\end{center}
\end{figure}

The fact that $S(x)$ is a finite band potential is important for getting self-consistency. At the same time, it
makes the problem analytically tractable. 
Of course, at this point one cannot be sure that the simplest finite band potential is sufficient to solve
the massive GN model at all temperatures and chemical potential, but this is what eventually will come out. 

\section{Minimizing the grand potential and self-consistency condition}

Having picked the ansatz for the scalar potential $S(x)$, we now have 
to minimize the grand potential with respect to the three parameters $A,\kappa, b$ and verify that it implies
self-consistency of the $\bar{\psi}\psi$-condensate. In HF
approximation, the
grand canonical potential density per flavor consists of the independent particle contribution $\Psi_1$ and
the double counting correction to the interaction $\Psi_2$,
\begin{eqnarray}
\Psi & = & \Psi_1 + \Psi_2 ,
\nonumber \\
 \Psi_1 &=& -\frac{1}{\beta\pi} \int_0^{\Lambda/2}\!\!\!\dd q \, \ln \left[\left(1+\ee^{-\beta(E-\mu)}\right)\left(1+\ee^{\beta (E+\mu)}
\right)\right], \nonumber \\
\Psi_2 & = & \frac{1}{2Ng^2 \ell}\int_0^\ell\!\!\dd x (S(x)-m_0)^2,  \qquad \ell= 2{\bf K}/A
\label{4.1}
\end{eqnarray}
(in the path integral approach, $\Psi_1$ and $\Psi_2$ correspond to the fermion determinant and the tree level term,
respectively).
$\Lambda/2$ is an UV cutoff which will eventually be sent to infinity, $\ell$ the spatial period of $S(x)$,
$E$ the single particle energy. 

In $\Psi_2$, we need the spatial averages of $S$ and $S^2$. 
It is convenient to introduce three basic functions of $b$ and $\kappa$, 
\begin{equation}
s=\frac{1}{{\rm sn}^2 b},  \qquad t=\frac{\cn\, b\, \dn\, b}{\sn^{3} b}. \qquad u=1-\EK  ,
\label{4.2}
\end{equation}
where $s$ and $t$ are related by
\begin{equation}
t^2= s(s-1)(s-\kappa^2).
\label{4.3}
\end{equation}
The spatial averages can then be written in the compact form
\begin{eqnarray}
\label{4.4}
\langle S\rangle & = & A\left(\mathrm{Z} + t/s\right) , \nonumber \\
\langle S^2 \rangle  & = & A^2 \left(s-1+2u -\kappa^2\right)
\end{eqnarray}
with Jacobi's Zeta function $\mathrm{Z}=\mathrm{Z}(b,\kappa)$. 

Turning to $\Psi_1$, let us first transform the momentum integral into an integral over Dirac energies. 
The density of states for the Lam\'e potential \cite{L28} implies a change of integration measure 
($p=q/A,\omega=E/A$)
\begin{eqnarray}
\frac{\dd p}{\dd \omega} &=&  \frac{\omega (\omega^2- s+u)}{\pm\sqrt{W}} ,
\nonumber \\
W &=& (\w^2- s+1)(\w^2- s+\kappa^2)(\w^2- s),
\label{4.5}
\end{eqnarray}
where the plus sign refers to the upper band, the minus sign to the lower band. 
With the shorthand notation
\begin{equation}
a=\beta A,\quad \nu=\mu\beta  ,
\label{4.6}
\end{equation}
$\Psi_1$ can be written as the following integral over the allowed bands
\begin{equation}
\pi\beta^2  \Psi_1  = -a\left( \int_{\sqrt{s-1}}^{\sqrt{s-\kappa^2}}\!\!\dd\w+\int_{\sqrt{s}}^{\Lambda_\w}\!\!\dd\w\right) 
\frac{\dd p}{\dd \w }
\ln\left[(1+\ee^{-a\w+\nu})(1+\ee^{a\w+\nu})\right] .
\label{4.7}
\end{equation}
The energy cutoff $\Lambda_\w$ has to be computed from the momentum cutoff $\Lambda/2$ and equation (\ref{4.5}),
\begin{equation}
\Lambda_\w = \frac{\Lambda}{2A} +\frac{A\langle \tilde{S}^2 \rangle}{\Lambda} + {\rm O}(\Lambda^{-3}) .
\label{4.8}
\end{equation}
Due to the quadratic divergence of the integral, it is necessary to keep the next-to-leading order term here.
Following references \cite{L27,L13b}, we combine the integral over both
energy bands
as well as over positive and negative energy modes into the real part of a line integral in the complex $\omega$-plane. The
path of integration runs infinitesimally above or below the real axis,  
\begin{equation}
\pi\beta^2  \Psi_1  = -a \lim_{\epsilon\to 0} \re \int_{-\Lambda_\omega +\ii \epsilon }^{\infty+\ii \epsilon} \dd\omega 
\frac{\omega(\omega^2- s+ u)}{\sqrt{W}} \ln\left(1+\ee^{-a\omega +\nu}\right) .
\label{4.9}
\end{equation}
We have to minimize $\Psi$ with respect to $A,b,\kappa$ for fixed $\mu,\beta,\gamma$, a rather tedious task
at first glance. In order to 
simplify the computations, we use the freedom to minimize $\Psi$ with respect to any other set of independent variables.
Specifically, we propose to replace $A,b,\kappa$ by the spatial
averages  $\langle S \rangle$, $\langle S^2 \rangle$ and the spatial period $\ell$ of $S$, see equations (\ref{4.1}) 
and (\ref{4.4}).
It turns out that these variables are extremely convenient for the proof of self-consistency.
The stationarity conditions for the grand potential then read
\begin{eqnarray}
\frac{\partial}{\partial \langle S \rangle} \pi\beta^2\Psi & = &  \beta F_0 \ =\ 0,
\nonumber \\
\frac{\partial}{\partial \langle S^2 \rangle}\pi\beta^2\Psi & = &  \beta^2 F_1/2 \ =\ 0,
\nonumber \\
\frac{\partial}{\partial \ell} \pi\beta^2\Psi & = & a^2 F_2/\ell \ = \ 0, 
\label{4.11}
\end{eqnarray} 
where we have introduced functions $F_i$ which play a key role in our approach.
The derivatives with respect to the new, composite variables can be taken trivially in the case of $\Psi_2$ since
\begin{equation}
\Psi_2 = \frac{1}{2Ng^2} \left( \langle S^2 \rangle - 2 m_0 \langle S \rangle + m_0^2 \right).
\label{4.12}
\end{equation}
Unfortunately this is not true for $\Psi_1$ available only in terms of the original variables $A,b,\kappa$ in equation
(\ref{4.9}). 
In order to compute $F_i$, we therefore invoke the chain rule,
\begin{equation}
\left(\begin{array}{c}
\partial_b \\ \partial_A \\ \partial_{\kappa}
\end{array} \right) \pi\beta^2\Psi = 
\left(\begin{array}{ccc}
\partial_b \langle S\rangle & \partial_b \langle S^2\rangle & \partial_b \ell \\
\partial_A \langle S \rangle & \partial_A  \langle S^2\rangle  & \partial_A \ell \\
\partial_{\kappa} \langle S\rangle & \partial_{\kappa} \langle S^2\rangle & \partial_{\kappa} \ell
\end{array}\right)
\left(\begin{array}{c}
\beta F_0\\ \beta^2 F_1/2\\  a^2 F_2 /\ell
\end{array}\right) .
\label{4.13}
\end{equation}
Upon evaluating and inverting the Jacobian matrix on the right hand side of equation (\ref{4.13}), the $F_i$ can be expressed
 in terms of directly computable derivatives of $\Psi$. After some algebra \cite{L13b} we arrive at 
\begin{eqnarray}
F_0 & =&  -\frac{\pi \beta m_0}{Ng^2}  - t \re \lim_{\varepsilon\to 0^+}\int_{-\infty+\ii\varepsilon}^{\infty+
\ii\varepsilon} \dd \w 
\left(\frac{\partial}{\partial \w} \frac{1}{\sqrt{W}}\right)\ln\left(1+\ee^{-a\omega +\nu}\right) , \nonumber \\
	F_1 & = & \frac{\pi }{Ng^2} - 1  + \frac{1}{a} \re \lim_{\varepsilon\to 0^+}
\int_{-\frac{\Lambda\beta}{2a}+\ii\varepsilon}^{\infty
+\ii\varepsilon}
 \dd \w \left(\frac{\partial}{\partial \w} \frac{\w^2}{\sqrt{W}} \right) \ln\left(1+\ee^{-a\omega +\nu}\right),  \nonumber \\
  F_2 & = & \frac{1}{a} \re \lim_{\varepsilon\to 0^+}\int_{-\infty+\ii\varepsilon}^{\infty+\ii\varepsilon} \dd \w 
\Biggl[ \frac{\w (\w^2-s+u)}
{\sqrt{W}} \nonumber \\
& &  -\frac{\partial}{\partial \w} \frac{t \mathrm{Z} - u \w^2 + (\w^2-s+1)(\w^2-s+\kappa^2)}{\sqrt{W}} \Biggr]  \ln\left(1+
\ee^{-a\omega +\nu}\right)  .
\label{4.14} 
\end{eqnarray} 
All three functions $F_i$ vanish at the minimum of the thermodynamic potential (\ref{4.1}).

Equations  (\ref{4.9}), (\ref{4.12}) and (\ref{4.14}) are the basis for our computations. They are not yet in a form suitable
for numerical
calculations since they still involve bare parameters and a cutoff, but the present form is illuminating with respect
to self-consistency. We therefore first turn to the issue of self-consistency.

The self-consistency condition for the scalar potential $S(x)$ was given in equation (\ref{3.4}).
For a single mode, $\bar{\psi}\psi$
reads \cite{L29} 
\begin{equation}
\bar{\psi}\psi = \frac{\omega \tilde{S}-t/\omega}{\omega^2-s+u}  .
\label{4.16}
\end{equation}
We transform the thermal expectation value of $\bar{\psi}\psi$ once again into a complex integral, 
\begin{eqnarray}
\langle \bar{\psi}\psi\rangle_{\mathrm{th}} &=  & \frac{1}{\pi} \int_0^{\Lambda/2}\!\!\dd q\,\bar{\psi}\psi \left(\frac{1}
{\ee^{\beta(E-\mu)}
+1}-\frac{1}{\ee^{-\beta(E+\mu)}+1}\right) \nonumber \\
& = & \frac{A}{\pi} \re \lim_{\varepsilon\to 0^+}\int_{-\frac{\Lambda\beta}{2a}+\ii\varepsilon}^{\infty+\ii\varepsilon} \!\!\dd \w\, 
\frac{\w^2\tilde{S} 
-t}{\sqrt{W}}\frac{1}{\ee^{a \omega -\nu}+1}  .
\label{4.17}
\end{eqnarray}
A partial integration (picking up a boundary term) then enables us to express the thermal expectation
value in terms of the functions $F_0,F_1$ 
introduced above,
\begin{equation}
\langle \bar{\psi}\psi\rangle_{\mathrm{th}} = \frac{S(x)}{\pi}\left(F_1-\frac{\pi}{Ng^2}\right) + \frac{F_0}
{\pi \beta}+\frac{m_0}{Ng^2} .
\label{4.18}
\end{equation}
In the minimum of the grand canonical potential, $F_0$ and $F_1$ vanish. This reduces equation (\ref{4.18}) to 
equation (\ref{3.4}), thereby establishing exact self-consistency of the scalar condensate.

We now comment on the practical procedure to determine the phase diagram and thermodynamics of the
massive GN model within this framework. First, we have to eliminate the bare parameters $Ng^2, m_0$ and the cutoff
 $\Lambda$
in the standard way. Since all UV divergences are due to vacuum effects, there are no new difficulties as
compared to the $T=0$ case. All we need is the vacuum gap equation
\begin{equation}
\frac{1}{Ng^2} = \frac{1}{\pi} (1+ m_0)\ln \Lambda = \frac{1}{\pi} (\ln \Lambda+ \gamma)  .
\label{4.19}
\end{equation}
The dynamical fermion mass in the vacuum (set equal to 1) and the confinement parameter $\gamma$  
are physical parameters which are kept fixed while one lets $Ng^2\to 0$, $m_0\to 0$ and
$\Lambda \to \infty$. Irrelevant divergent terms $-\Lambda^2/8\pi$ and 
$-\mu\Lambda/2\pi$ from
the energy and baryon density of the Dirac sea can simply be dropped. An expression for the renormalized
grand canonical potential in which the limit $\Lambda\to \infty$ can safely be taken is
\begin{eqnarray}
\pi\beta^2\Psi_{\rm ren} &=&  \lim_{\Lambda\to\infty} \Biggl[ \frac{\Lambda^2\beta^2}{8} +\frac{a^2 \langle 
\tilde{S}^2\rangle}{2}\left(\ln 
\Lambda +\gamma -1 \right) - a\beta\gamma \langle\tilde{S}\rangle \nonumber \\
&- &  a \re \lim_{\varepsilon\to 0^+}\int_{-\frac{\Lambda\beta}{2a}+\ii\varepsilon}^{\frac{\Lambda\beta}{2a}+\ii\varepsilon} \dd \w
 \frac{\omega(\omega^2- s+u)}{\sqrt{W}} \ln \left(2\cosh \frac{a\w -\nu}{2}\right) \Biggr]  .
\label{4.20}
\end{eqnarray}

In principle, we have to solve the three equations $F_i=0$ simultaneously. 
It is worthwhile to examine more closely the way in which the $F_i$  depend on the 6 relevant parameters
($a,b,\kappa,\gamma, \nu,\beta$), since this suggests a simpler strategy of how to minimize $\Psi$. 
We first note that $F_2$ is a convergent integral which does not require any regularization or renormalization.
Since $F_2$ depends neither on $\gamma$ nor on $\beta$, the equation 
\begin{equation}
F_2(a,b,\kappa,\nu)=0
\label{4.21}
\end{equation}
can be solved for $a$, say, for given $b,\kappa,\nu$. Now let us focus on the $\gamma, \beta$ dependence
of the other two equations.
$F_0$ is also free of divergences. All we have to do here is to 
replace the ratio of bare parameters $\pi m_0/Ng^2$ by $\gamma$. The equation $F_0=0$ is then turned into
\begin{equation}
\gamma \beta = a t I_0(a,b,\kappa,\nu), 
\label{4.22}
\end{equation}
where $I_0$ can be inferred from equation (\ref{4.14}). 
The integral in $F_1$ on the other hand has a logarithmic divergence at the lower integration limit.
Isolating the divergence and eliminating $1/Ng^2$ with the help of the gap equation (\ref{4.19}), the ln $\Lambda$ terms
are cancelled and the equation $F_1=0$ assumes the form 
\begin{equation}
\gamma - \ln \beta  = I_1(a,b,\kappa,\nu).
\label{4.23}
\end{equation}
Equations  (\ref{4.22}) and (\ref{4.23}) can be combined into 
\begin{eqnarray}
\gamma + \ln \gamma &=& I_1 + \ln \left( a t I_0 \right),
\nonumber \\
\beta & = & a t I_0 /\gamma.
\label{4.24}
\end{eqnarray}
The first of these equations can be solved for $\gamma$, the second one then yields an explicit expression for $\beta$.
In total we have reduced the problem of finding the minimum of a function of three variables to the simpler problem of 
finding the zero's of two functions of one variable each. 

Throughout this section, we have made use of complex integration so as
to exhibit the formal structure of various expressions and the self-consistency in the most transparent way.
In order to actually compute the one-dimensional numerical integrals in $\Psi$ and in
the $F_i$, it is advisable to convert the integrals back to standard real integrals
over the allowed energy bands. For more technical details and explicit expressions for $I_0,I_1,F_2$ we refer the reader
to reference \cite{L13b}. 

\section{The full phase diagram of the Gross-Neveu model}
In the previous section, we have outlined how to minimize the grand canonical potential given the 3-parameter ansatz for
$S(x)$. This can be used to evaluate bulk thermodynamic observables for any given $T,\mu$ and $\gamma$.
In the present section we turn to the structure of the phase diagram, including
the location of the phase boundaries, the order of
the transitions and the symmetries of the various phases. The following considerations 
have proven useful for this purpose.
We expect phase transitions between a crystal phase and a homogeneous phase where $S(x)=M$ is constant.
In the chiral limit, we can further distinguish between a chirally restored phase ($M=0$) and a massive, chirally
broken phase ($M>0$). We therefore first need to understand for which choice of parameters $S(x)$, equations
(\ref{3.9},\ref{3.13}), goes over into a constant.
The elliptic modulus $\kappa$ varies between 0 and 1. At the two boundaries of this interval, we find
\begin{eqnarray}
\lim_{\kappa \to 0} \tilde{S} &=& \cot b,
\nonumber \\
\lim_{\kappa \to 1 } \tilde{S} & = & \coth b + \tanh \xi - \tanh (\xi+b).
\label{5.1}
\end{eqnarray}
The first limit is a constant, the second describes the single baryon profile. Clearly, both limits are relevant for 
phase boundaries where $\tilde{S}$ goes over from a periodic function in the crystal phase to 
a constant one in the massive Fermi gas phase ($M=A \cot b$ or $A \coth b$, respectively).
As far as $b$ is concerned, important special values are $b=0$ and $b={\bf K}$ with
\begin{eqnarray}
\lim_{b\to 0} b \tilde{S} &=& 1,
\nonumber \\
\lim_{b \to {\bf K}} \tilde{S} &=& \kappa^2 \frac{{\rm sn}\, \xi \,{\rm cn}\, \xi}{{\rm dn}\, \xi}. 
\label{5.2}
\end{eqnarray}
The value $b=0$ is the only value of $b$ where $\kappa=0$ and $\kappa=1$ can coexist and
therefore plays a prominent role in the phase diagram. In the chiral limit, $b$ takes on the value ${\bf K}$.
If we let $b\to {\bf K}$ and $\kappa \to 0$ simultaneously, $\tilde{S}$ vanishes and
we can connect the crystal phase continuously to a chirally restored phase. 

These observations are the key for computing the phase boundaries separating the crystal phase from the 
Fermi gas. 
Evidently, they are only applicable if the phase transitions are continuous, but this is indeed what
we find.
For $\kappa=0$ or 1, the functions $F_i$ further simplify, so that the actual computation along the lines
described in section 4 is quite manageable.   
We have computed lines of constant $b$ and constant $\nu$, since this is most easily done in our scheme.
For each value of $\kappa$, a certain ($b,\nu$)-grid is mapped onto a two-dimensional surface in ($\gamma,\mu,T$)-space
by minimizing the grand potential. The resulting curved surface represents a 2nd order phase boundary. 

\begin{figure}[ht]
\begin{center}
\epsfig{file=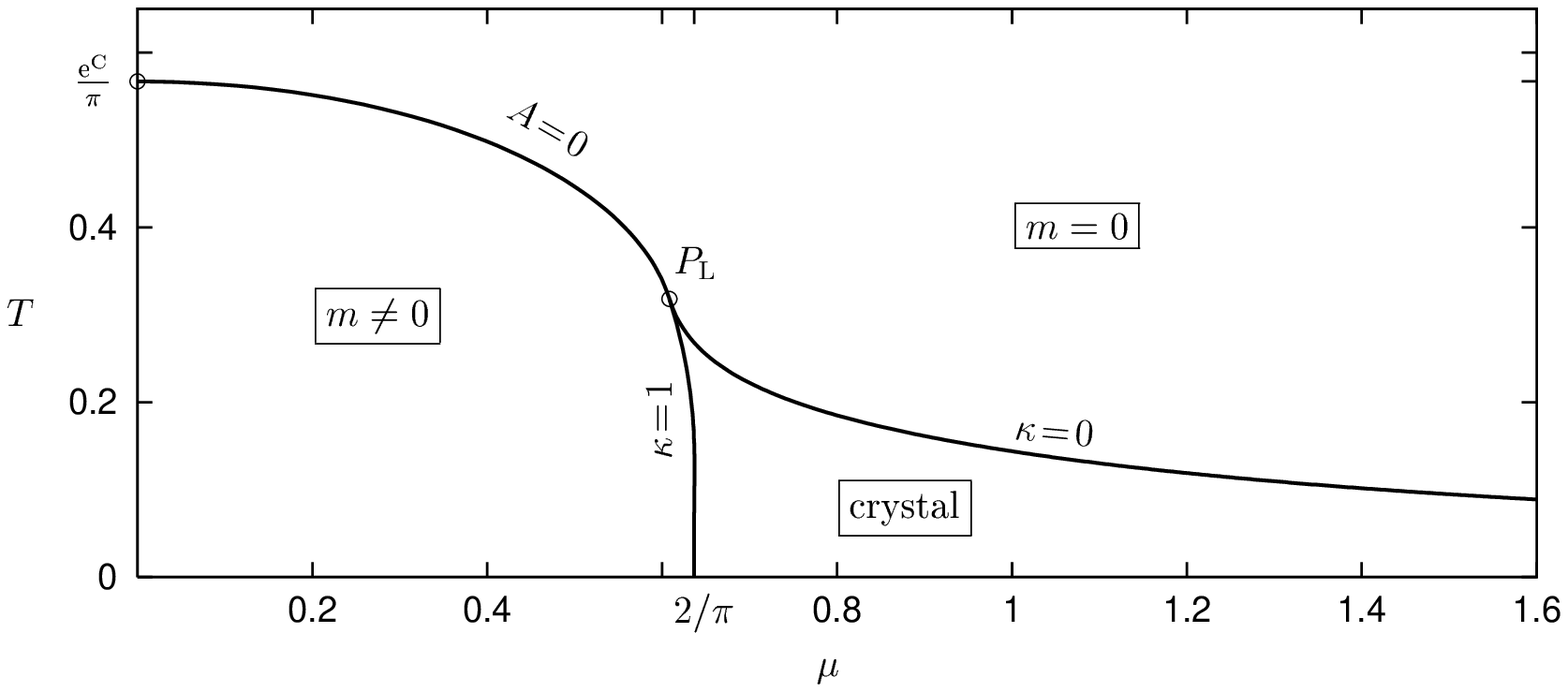,width=14.0cm}
\caption{Phase diagram of the GN model in the chiral limit \cite{L27,L30}.
All phase boundaries correspond to second order transitions.}
\end{center}
\end{figure}

We now turn to the numerical results, starting with the chiral limit $\gamma=0$ and focusing on the phase
boundaries. These are depicted in figure 4
which supersedes figure 1. The 2nd order line AB of the old phase diagram is unaffected. The first order line
BD of figure 1 is replaced by two 2nd order lines delimiting a novel kink-antikink crystal phase. The
tricritical point B is turned into another kind of multicritical point labeled $P_{\rm L}$ in figure 4, located
at precisely the same ($\mu,T$) values. As we switch 
on $\gamma$ (figure 5),
the second order line separating
massive ($M>0$) and massless ($M=0$) phases disappears as a consequence of the explicit breaking of chiral
symmetry. The crystal phase survives at all values of $\gamma$, but is confined
to decreasing temperatures with increasing $\gamma$. For fixed $\gamma$, it is bounded by two 2nd order lines joining in a
cusp. The cusp coincides with the critical point of the old phase diagram but has once again a significantly different
character. The crystal phase exists and is thermodynamically stable inside the tent-like structure formed out of two sheets
denoted as $I$ and $II$. These sheets are defined
by $\kappa=0$ ($I$) and $\kappa=1$ ($II$), respectively. The line P$_{\rm t}$C where they join corresponds to $b=0$
and coincides with line P$_{\rm t}$C  in figure 2. The baseline of sheet $II$ in the ($\mu,\gamma$)-plane
has a simple physical interpretation: It reflects the $\gamma$-dependence of the baryon mass in the massive GN
model, or equivalently the critical chemical potential at $T=0$.
The chiral limit $\gamma \to 0$ can be identified with $b \to {\bf K}$. 

Another view of the phase diagram is displayed in figure 6. Here we plot the phase boundaries in the
($\mu,T$)-plane for several values of $\gamma$. These two-dimensional
graphs correspond to cutting the three-dimensional graph in figure 5 by (equidistant) planes $\gamma=$ constant. 
They provide a better view of how the two critical lines are joined in a cusp and exhibit that the region in which
the crystal is thermodynamically stable shrinks with increasing bare fermion mass.

The cusp line where the sheets $I$ and $II$ are glued together can be determined by taking the limit $b\to 0$ in $F_i$.
As discussed above, in this limit the scalar potential becomes both homogeneous and $\kappa$-independent.
One can show analytically that the resulting equations $F_i=0$ for $i=0,1,2$ are equivalent to the conditions
$\Psi'=0,\Psi''=0,\Psi'''=0$ ($'=\partial_M$) for the translationally invariant calculation \cite{L12}. Hence 
the curve $b=0$ in the new phase diagram coincides with the line of critical points of the old phase diagram.
One can also convince oneself that the critical first order
phase transition sheet shown in figure 2 is tangential to both sheets $I$ and $II$ along the line of endpoints.

\begin{figure}[ht]
\begin{center}
\epsfig{file=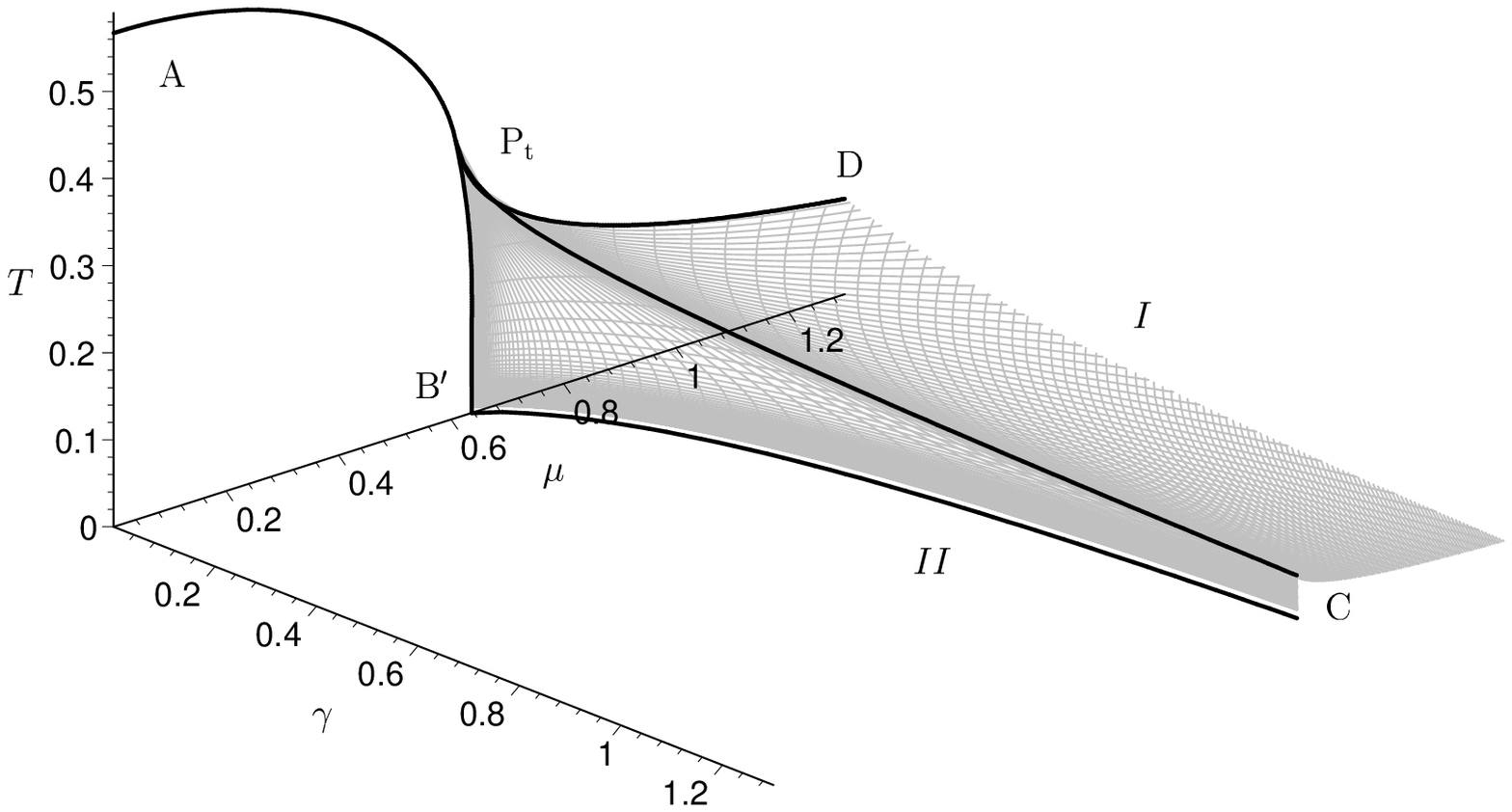,width=8cm,angle=270}
\vskip 1.0cm
\caption{Revised phase diagram of the massive Gross-Neveu model \cite{L13a,L13b}. The shaded surfaces $I,II$ separate
the kink-antikink
crystal from a massive Fermi gas and correspond to 2nd order phase transitions. The $\gamma=0$ plane is the same
as figure 4.}
\end{center}
\end{figure}

\begin{figure}[ht]
\begin{center}
\epsfig{file=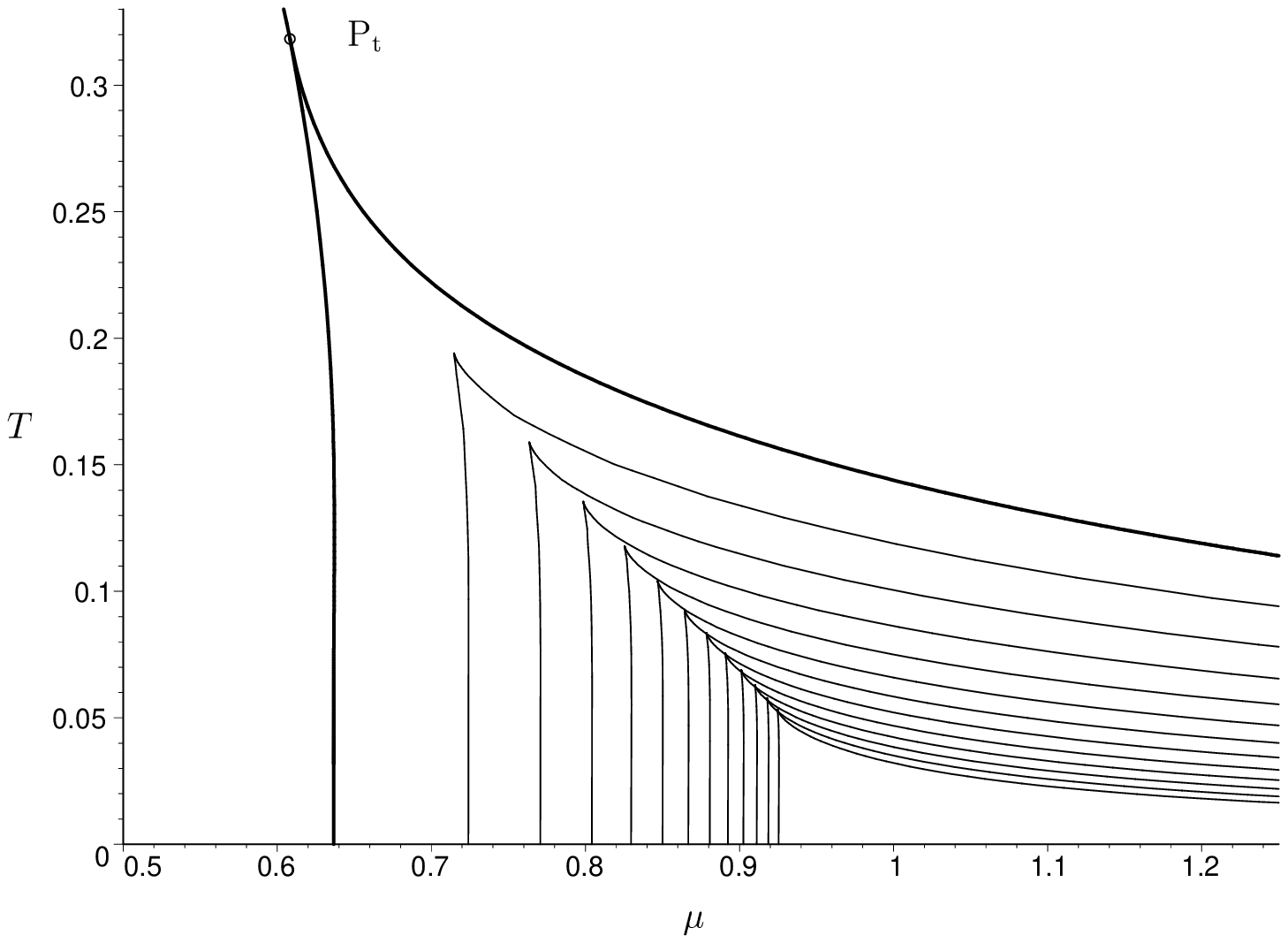,width=8cm,angle=270}
\vskip 1.0cm
\caption{Two-dimensional sections $\gamma=$ const. through the phase diagram of figure 5. 
The fat lines belong
to $\gamma=0$, the thin lines to $\gamma=0.1,0.2,...,1.2$, from top to bottom. The position of the cusp agrees with 
the critical point in the old phase diagram. From reference \cite{L13b}.}
\end{center}
\end{figure}
\section{Ginzburg-Landau theory near the tricritical point}

In the vicinity of the tricritical point P$_{\rm t}$, we can derive a Ginzburg-Landau
effective action by expanding $\Psi_{\rm ren}$ in $a=\beta A$ \cite{L27,L13b}. Since $S(x)$ is both weak and slowly varying
there, this could also be obtained approximately without knowing the exact solution. Here, we turn things
around and derive it via a Taylor expansion from the full expression. This should help to understand the character of
the multicritical point.
   
We therefore start with an expansion of $\Psi_{\rm ren}$ in equation (\ref{4.20}) in powers of $a$,
\begin{eqnarray}
\pi\beta^2\Psi_{\rm ren}  &=& -\frac{\nu^2}{2} - \frac{\pi^2}{6} -\gamma\beta^2 \langle S \rangle - \frac{\beta^2}{2} \langle
 S^2\rangle \left(\ln\frac{\beta}{4\pi}-\gamma\right) \label{6.1}  \\
& & + a^2\sum_{n=0}^\infty \left(-\frac{a^2}{4\pi^2}\right)^n \frac{c_{n+1}+(u-s)c_n}{(2n+1)!}\re \, \psi \! \left( 2n , 
\frac{1}{2} + \frac{\ii\mu}{2\pi T}\right).
\nonumber 
\end{eqnarray}
Here, $\psi(n,z)$ is the polygamma function, the $n$-th derivative of the digamma function $\psi(z)=\Gamma'(z)/\Gamma(z)$.
The coefficients $c_n$ can be obtained from a generating function,
\begin{equation}
\frac{\w^3}{\sqrt{W}} =  \sum_{n=0}^\infty \frac{c_n}{\w^{2n}} .
\label{6.2}
\end{equation}
One can now verify that the combinations $c_{n+1}+(u-s)c_n$ appearing in equation (\ref{6.1}) are related as follows
to spatial averages of powers of $S$ and its derivatives with respect to $x$,
\begin{eqnarray}
c_1+ (u-s) c_0 & = &  \frac{1}{2A^2} \langle S^2 \rangle  \nonumber \\
c_2 + (u-s) c_1 & = & \frac{3}{8A^4}\left(\langle S^4 \rangle +\langle (S')^2\rangle \right) \nonumber \\
c_3+ (u-s) c_2 & = & \frac{5}{16A^6} \left(\langle S^6\rangle +\frac{1}{2}\langle(S'')^2\rangle+
5\langle S^2(S')^2 \rangle \right)  .
\label{6.3}
\end{eqnarray}
This enables us to write down the Ginzburg-Landau effective action in the form
\begin{equation}
\Psi_{\rm eff}(\gamma)  = \Psi_{\rm eff}(\gamma=0) +\frac{\gamma}{2\pi}\left(S^2 - 2S\right),
\label{6.4}
\end{equation} 
where the effective action at $\gamma=0$ is 
\begin{eqnarray}
\fl
  \Psi_{\rm eff}(\gamma=0) & =&  - \frac{\pi}{6} T^2 - \frac{\mu^2}{2\pi} + \frac{1}{2\pi}  S^2  \left[ \ln (4\pi T)+ 
{\rm Re}\  \psi \left( \frac{1}{2} +
 \frac{\ii\mu}{2\pi T}\right) \right] \nonumber \\
\fl
  && - \frac{1}{2^6 \pi^3 T^2} \left(  S^4  +  (S')^2  \right) \re \  \psi \left( 2, \frac{1}{2} + \frac{\ii\mu}{2\pi T}\right) \nonumber \\
\fl
&  & + \frac{1}{2^{11}3\pi^5 T^4} \left(  S^6  + \frac{1}{2}  (S'')^2 + 5  S^2 (S')^2  \right) \re\  \psi \left( 4, \frac{1}{2} +
 \frac{\ii\mu}{2\pi T}\right) .
\label{6.5}
\end{eqnarray}
The instability with respect to crystallization is related to the fact that the ``kinetic" term $\sim (S')^2$
can change sign, depending on $\mu$ and $T$. 
We will come back to this formula in section 11 where we point out that it has a literal correspondence in condensed matter
physics, namely in the theory of inhomogeneous superconductors.

Notice that if one drops all derivatives of $S$ in the effective action, one gets the result for the GN model
under the assumption of unbroken translational invariance. The tricritical point for instance is defined
by the simultaneous vanishing of the $S^2$ and $S^4$ coefficients which happens at $\mu_t=0.608221$,
$T_t= 0.318329$. Since the coefficients are the same as in the full effective action, we 
can understand why the tricritical point stays at the same place even though we allow for $x$-dependent potentials. 

\section{More about the chiral limit}

In references \cite{L27,L30}, a number of additional results has been obtained in the chiral limit $\gamma=0$. Here
 we give a selection of
figures and formulas to highlight certain features of this simpler special case.

We first reiterate that only at $\gamma=0$ we are dealing with three distinct phases: the crystal, a massive and a massless 
Fermi gas. The scalar potential can be obtained by letting $b \to {\bf K}$ in equations (\ref{3.9},\ref{3.13}),
\begin{equation}
\tilde{S}(\xi) = \kappa^2 \frac{{\rm sn}\, \xi \,{\rm cn}\, \xi}{{\rm dn}\, \xi} = \kappa^2 {\rm sn}\, \xi \, {\rm sn}(\xi+{\bf K})
\label{7.1a}
\end{equation}
and has a higher symmetry than for $\gamma \neq 0$, namely
\begin{equation}
S(x+\ell/2)=-S(x)
\label{7.1b}
\end{equation}
where $\ell$ is the spatial period.
This is actually a remnant of the original discrete chiral symmetry of the model. Translational invariance and the $\gamma^5$
transformation both break down, leaving the unbroken discrete symmetry (\ref{7.1b}), i.e., a translation by half a period
combined with a $\gamma^5$ transformation.

An expanded plot which reveals more details about the shape of the phase boundary
separating crystal and massive Fermi gas (labeled $\kappa=1$) is displayed in figure 7. 

\begin{figure}[ht]
\begin{center}
\epsfig{file=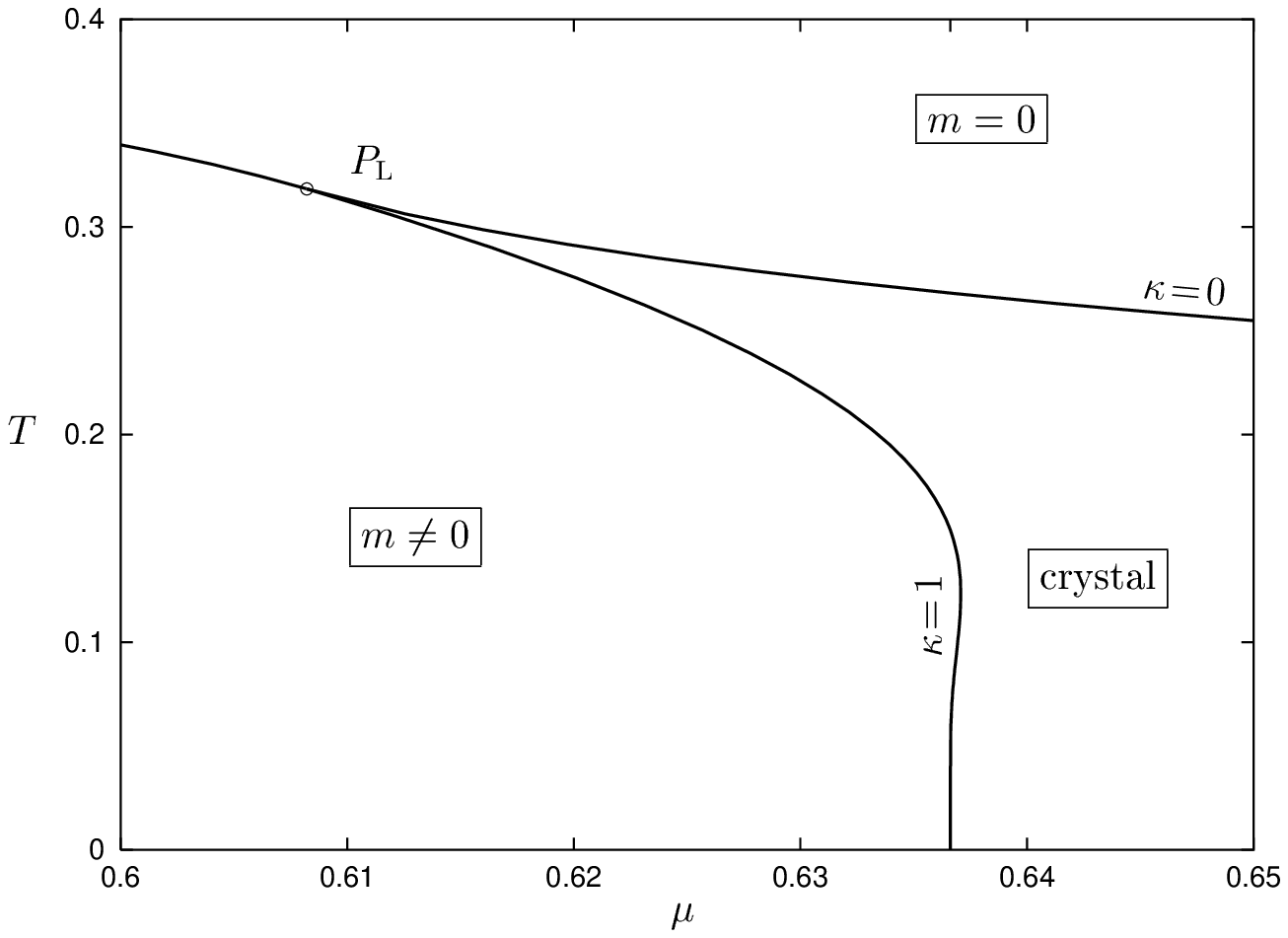,width=10cm}
\caption{Close-up on the phase boundary between homogeneous and inhomogeneous ordered phases ($\kappa=1$).
Notice the different scale on the $\mu$-axis as compared to figure 4. From reference \cite{L27}.}
\end{center}
\end{figure}

\begin{figure}[ht]
\begin{center}
\epsfig{file=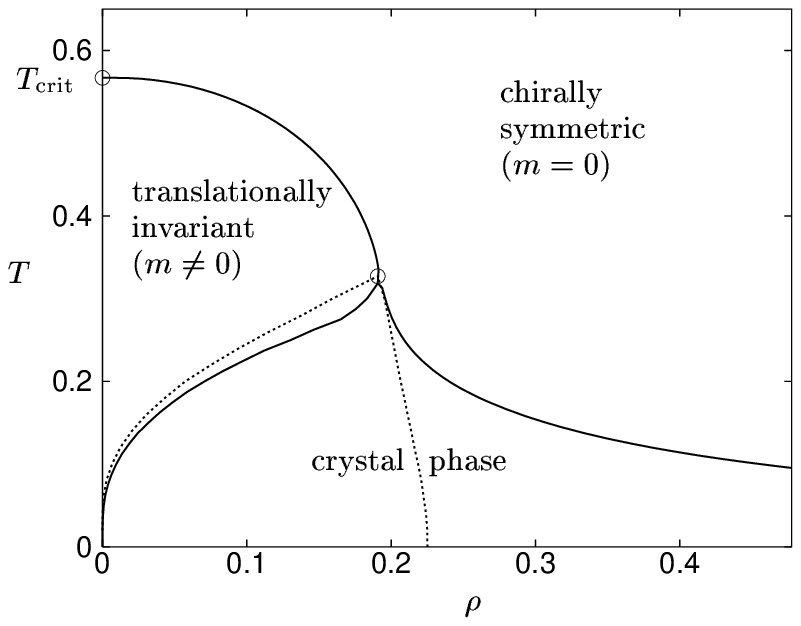, width=10cm}
\end{center}
\caption{Revised phase diagram of the GN model in the ($T,\rho$)-plane \cite{L30}. The dashed lines 
belong to the old phase diagram, where they enclose the mixed phase. This ``droplet" region
is superseded by the crystal phase featuring baryons.}
\end{figure}
A way of presenting the phase diagram complementary to that in figure 4 is given in figure 8
where we have transformed all phase boundaries from the ($T,\mu$)- into the ($T,\rho$)-plane. 
Here, the first order line of the old solution splits up into the two dashed lines which delimit the mixed phase
region (droplets of chirally restored matter in the chirally broken vacuum). This
should be replaced now by the two solid lines going downward from the tricritical point and
enclosing the crystal phase. At $T=0$ in particular, the crystal phase is stable at all densities.

As we approach the $\kappa=0$ phase boundary, $S \sim \cos 2qx$ (with an amplitude vanishing at $\kappa=0$), and 
the wave number $q$
can serve as order parameter for the breakdown of translational invariance.
In figure 9 we show the dependence of this order parameter on $\mu$ as one moves along
the phase boundary.
The solid line is the curve $q=\mu$ which is approached asymptotically by the full calculation.  
At $\mu=\mu_t$, the tricritical point of the old solution, we see a clear signal of a 2nd order phase transition with breakdown
of translational invariance. 

\begin{figure}[ht]
\begin{center}
\epsfig{file=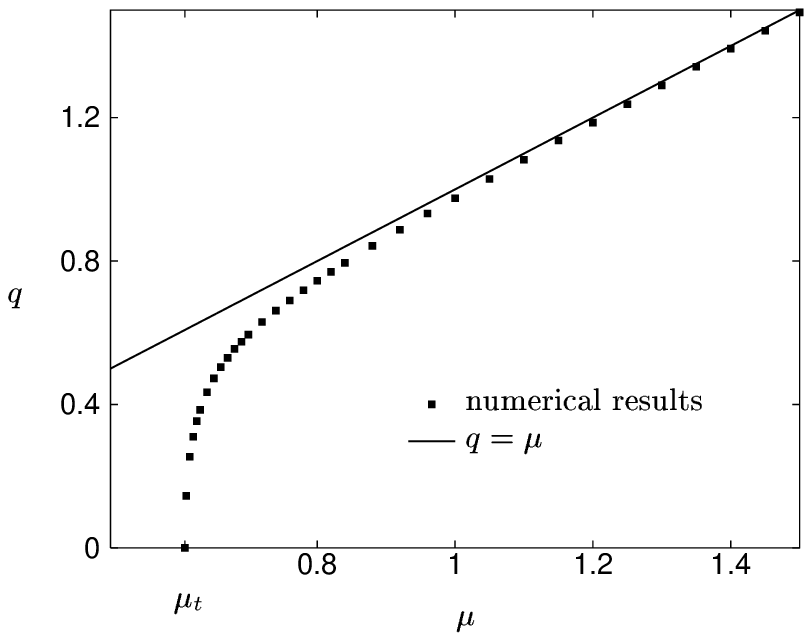, width=10cm}
\end{center}
\caption{Wave number characterizing the crystal period along the phase boundary, showing 
a continuous phase transition at the tricritical point $\mu_t$. From reference \cite{L30}.}
\end{figure}

Since the phase boundaries are of particular interest and easier to compute than other thermodynamic
quantities, let us  
give the explicit formulas derived in reference \cite{L27} from which they can be obtained.

The boundary between the chirally restored phase and the crystal is characterized by $\kappa=0$. It can either
be derived by almost degenerate perturbation theory \cite{L30} or from the full thermodynamic
potential. In the latter case, a straightforward expansion around $\kappa=0$ yields
\begin{equation}
\label{7.2}
\ln\frac{1}{4\pi T}=\frac{1}{2}\min\limits_{a\geq 0}\mathrm{Re}\left[\psi\left(\frac{1}{2}+\frac{\mathrm{i}}{2\pi}(\nu+a)\right) + 
(a\rightarrow -a)\right]\, .
\end{equation}
The resulting curve is labeled ``$\kappa=0$" in figures 4 and 7. 
The asymptotic behavior for large $\mu$ can also be determined,
\begin{equation}
T_{\rm crit} = \frac{{\rm e}^{\rm C}}{4\pi \mu}  .
\label{7.3}
\end{equation}
The phase boundary does not reach the $T=0$ axis so that chiral symmetry is not restored at $T=0$ no matter how high 
the density is, at variance with the naive expectation for an asymptotically free theory.

For small $\nu$ where $\mathrm{Re} \left[\psi\left(2,\frac{1}{2}+
\frac{\mathrm{i}\nu}{2\pi}\right)\right]<0$, the unique minimum is
at $a=0$. In this
range of $\nu$ the phase boundary does not touch the crystal region in the $(\mu,T)$-diagram. It corresponds 
to the transition 
between the massless and the massive homogeneous solutions described by
\begin{equation}
\label{7.4}
\ln\frac{1}{4\pi T} = \mathrm{Re} \, \psi\left(\frac{1}{2}+\frac{\mathrm{i\nu}}{2\pi}\right)\, .
\end{equation}
Upon using $\psi(1/2)=- \mathrm{C}-\ln4$ ($\mathrm{C}\approx 0.5772$ is the Euler constant), we reproduce the well-known
value of the critical temperature at $\mu=0$,
\begin{equation}
T_c=\frac{\mathrm{e}^{\mathrm{C}}}{\pi} \, .
\label{7.5}
\end{equation}

Next we turn to the non-perturbative phase boundary at $\kappa \to 1$. It separates
the crystal from the massive phase. 
The relation between $a$ and $\nu$ along the phase boundary can be deduced from 
\begin{equation}
\label{7.7}
\frac{\partial}{\partial a}\frac{1}
{a}\int\limits_0^{\pi/2}
\mathrm{d}\varphi \frac{1}{\cos\varphi} \mathrm{Im} \ln \frac{\Gamma\left(\frac{1}{2}+\frac{i}{2\pi}(\nu+a\cos \varphi)\right)}
{\Gamma\left(\frac{1}{2}+\frac{i}{2\pi}(\nu-a\cos \varphi)\right)} =0  .
\end{equation}
With these values for $a$ and $\nu$,
\begin{equation}
\label{7.8}
-\ln\frac{\beta}{4\pi}+\frac{1}{\pi}\int_0^\pi\mathrm{d}\varphi \, \mathrm{Re}\, \psi\left(\frac{1}{2}+
\frac{\mathrm{i}}{2\pi}(\nu
+a\cos\varphi)\right) = 0
\end{equation}
yields $\beta$ along the phase boundary. The resulting curve is shown in the phase diagram in figures 4 and 7
(label ``$\kappa=1$"). Equation (\ref{7.8}) alone with 
values of $(a,\nu)$
that are not restricted by equation (\ref{7.7}) gives the connection between $a$ (or, equivalently, the effective fermion mass),
$\nu$ and $\beta$ in the massive phase outside the crystal region.

\section{More about the zero temperature limit}

Let us begin with some additional plots on cold, dense matter in the massive GN model \cite{L29}.
First of all we wish to illustrate how the self-consistent scalar potential depends on $\gamma$. This is exhibited
in figure 10 at low and high density, respectively.
\begin{figure}[ht]
\begin{center}
\epsfig{file=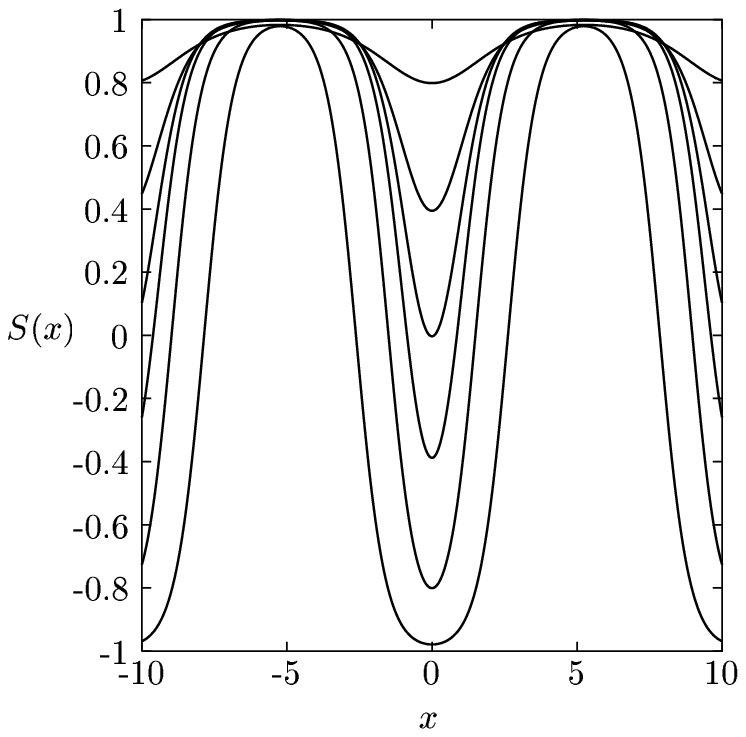,width=8cm}\epsfig{file=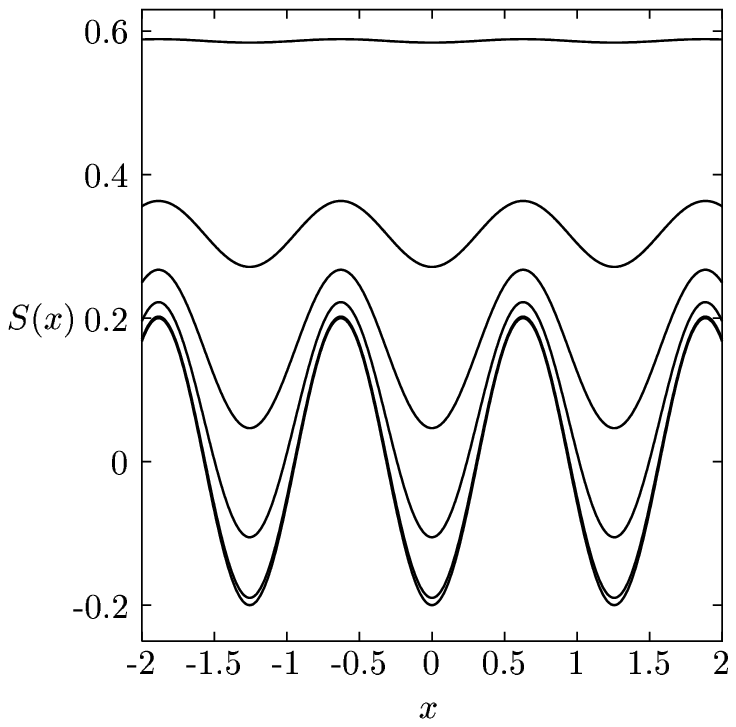,width=8cm}
\caption{Self-consistent scalar potential $S(x)$ versus $x$ for $p_f=0.3$ (left) and $p_f=2.5$ (right). From bottom to top:
$\gamma= 0,0.01,0.1,0.3,0.75,2.3$. From reference \cite{L29}.}
\end{center}
\end{figure}
The deepest curves always correspond to $\gamma=0$, where the potential oscillates symmetrically around zero
due to the residual symmetry (\ref{7.1b}).
As we increase the symmetry violation parameter $\gamma$, the potential oscillates with 
decreasing amplitude around a value close to the mass $M$ which the fermions would acquire 
in the translationally invariant
solution.
It is surprising that such a variety of potential shapes in the Dirac equation can all be reduced to the 
standard single gap Lam\'e equation.

The next result which we should like to show is how the density varies with the chemical potential. 
At $T=0$, the chemical 
potential can be obtained by differentiating the energy density with respect to the mean fermion density,
\begin{equation}
\mu =   \frac{\partial E_{\rm g.s.}}{\partial \rho}, \qquad \rho=\frac{p_f}{\pi}.
\label{8.1}
\end{equation} 
If we assume unbroken translational invariance (figure 11, left), we find discontinuities in these curves, confirming the result
of reference \cite{L12} about a first order phase transition. Repeating the same calculation for the crystal solution (which
is the
stable one), all the curves become continuous, signaling a 2nd order phase transition (figure 11, right).   
The critical chemical potential in this latter case coincides with the baryon mass, as expected on general grounds.
By contrast, the first order transition in figure 6 happens at a chemical potential which has at best the meaning of an
approximate baryon mass in a kind of droplet model, cf. the discussion in section 2. 
\begin{figure}[ht]
\begin{center}
\epsfig{file=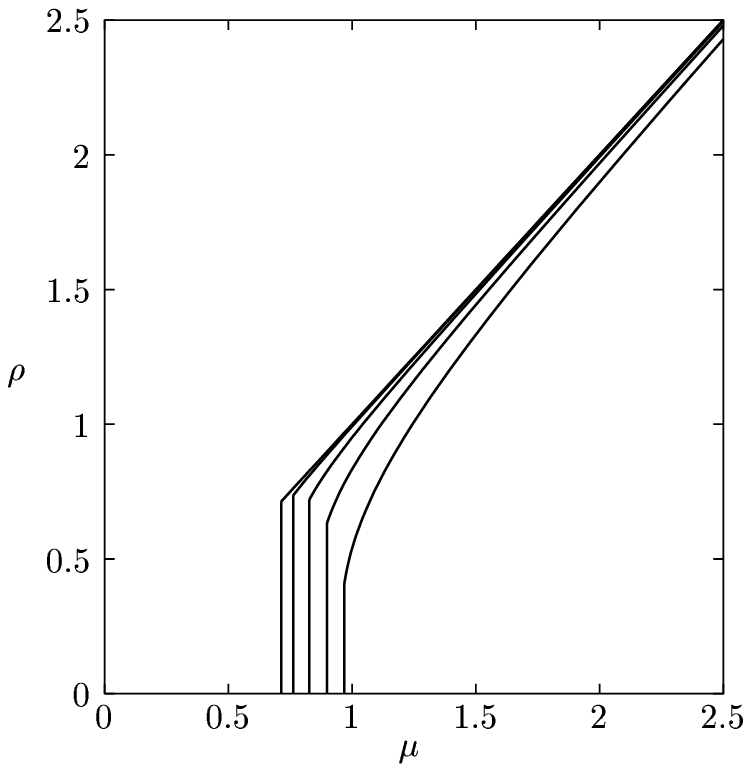,width=8cm}\epsfig{file=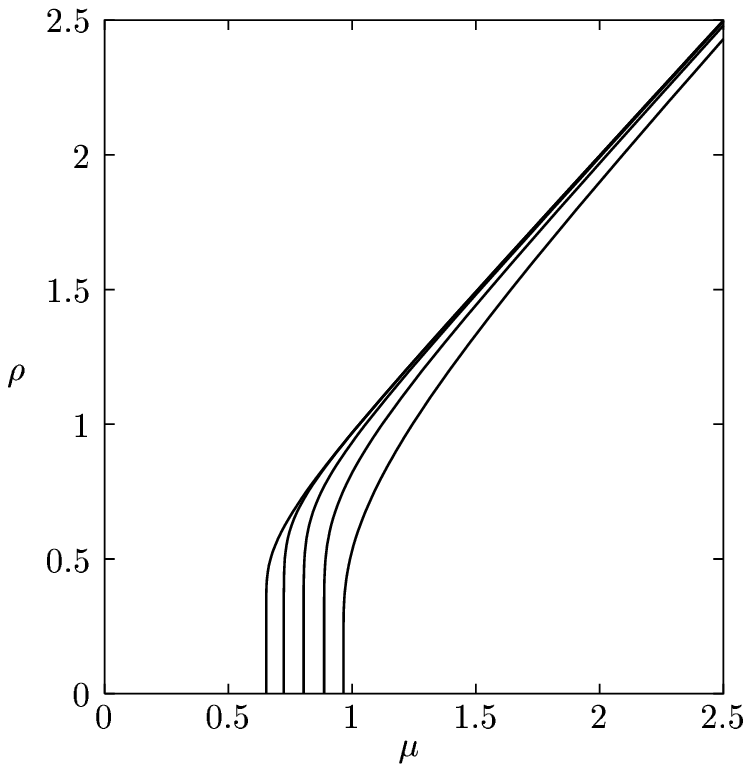,width=8cm}
\caption{Baryon density versus chemical potential \cite{L29}. Left: (unstable) translationally invariant solution, showing
a first order transition.
Right: (stable) crystal solution with 2nd order transition. Curves from left to right: $\gamma=0.01,0.1,0.3,0.75,2.3$.}
\end{center}
\end{figure}

The two real parameters which determine $S(x)$ at $T=0$ are the elliptic modulus $\kappa$ and the shift $b$.  
The scale factor $A$ on the other hand is determined by the baryon density and $\kappa$, 
\begin{equation}
A= \frac{2 p_f {\bf K}}{\pi}.
\label{8.2}
\end{equation}
Expression (\ref{8.2}) implies that the mean density scales with the inverse spatial period of the potential, 
a consequence of the fact that the valence band is completely filled (for matter) or empty 
(for antimatter). 

We now turn to additional analytic results. At $T=0$, all integrals can be reduced to incomplete elliptic integrals
of first, second and third type $F,E,\Pi$. We illustrate this fact in the case of the
ground state energy and the self-consistency conditions.
The ground state energy density is split up according to $E_{\rm g.s.}=E_1+E_2$ where $E_1$ is the sum over
single particle energies
of the occupied states, $E_2$ the double counting correction to the interaction energy. For $E_1$ we get
\begin{eqnarray}
\fl
\frac{2\pi}{A^2} E_1 & = &  \left( 2 \frac{\bf E}{\bf K} - 2 + \kappa^2- \frac{1}{2} \eta\right)
 +  \left( 2 \frac{\bf E}{\bf K} - 2 -\eta\right) \ln \frac{\Lambda}{A\sqrt{2+\eta}} +    \chi E(\tilde{p},q) 
\label{8.3} \\
\fl
& & +   \frac{\kappa^2}{\chi}\left(2\frac{\bf E}{\bf K} -2 - \eta \right) \left[F(\tilde{p},q)
-\Pi\left(\tilde{p},(\kappa')^2,q\right)\right]
 +   \chi \left( 2 \frac{\bf E}{\bf K} - 2 + \kappa^2\right) F(\tilde{p},q).
\nonumber
\end{eqnarray}
with the following definitions [$\eta$ has been defined in equation (\ref{3.15})],
\begin{eqnarray}
\chi & = & \sqrt{1+\eta},
\nonumber \\
\tilde{p} & = & {\rm dn}\, b,
\nonumber \\
q & = & \kappa'/{\rm dn}\, b.
\label{8.3a}
\end{eqnarray}
$E_2$ on the other hand coincides with $\Psi_2$ in equation (\ref{4.1}) since the HF double counting correction 
is $T$-independent.
After renormalization and upon using the vacuum gap equation together with equations (\ref{4.4}), one finds
\begin{equation}
E_2 = \frac{A^2}{2\pi}\left( \gamma +  \ln \Lambda \right) (s-1+2u-\kappa^2)
-  \frac{A}{\pi}\gamma    (\mathrm{Z}+t/s). 
\label{8.5}
\end{equation}
When adding up $E_1$ and $E_2$ the logarithmic divergence drops out and a finite result depending only on physical
parameters is obtained.

Consider the self-consistency condition next.
For $\gamma \neq 0$, it can be cast into the form
\begin{eqnarray}
0 & = & A\, {\rm cn}\, b \, F(\tilde{p},q) -  \gamma\,  {\rm sn}^2 b,
\label{8.8} \\
\gamma & = & A \, {\rm cn}\, b  \left[ \kappa^2 \Pi\left(\tilde{p},(\kappa')^2,q\right) + \chi
\left( \gamma  + \ln (A \sqrt{2+\eta}) \right) \right].
\nonumber
\end{eqnarray}
The original $x$-dependent condition has been transformed into two $x$-independent equations.
They determine $\kappa$ and $b$ for given $p_f$ and $\gamma$.

We finish this section with a few words about the high density limit where the calculation 
becomes perturbative (see \cite{L29}). 
We use almost degenerate perturbation theory, following the standard weak coupling approach from
solid state physics textbooks. In the double-counting correction, we have to
take into account the bare fermion mass, and we allow for $S_0 \neq 0$ in addition to $S_{\pm 1}\neq 0$
[$S_{\ell}$ are the Fourier components of the periodic potential $S(x)$].
Thus our present ansatz for $S(x)$ is 
\begin{equation}
S(x)=S_0 + 2 S_1\cos ( 2p_f x) .
\label{8.9}
\end{equation}
The approximate energy density becomes   
\begin{equation}
\fl
E_{\rm g.s.}  =  - \frac{\Lambda^2}{8\pi} + \frac{p_f^2}{2\pi} + \frac{S_0^2}{2\pi} \ln (2p_f)
+ \frac{\gamma}{2\pi} \left( S_0^2-2 S_0 \right)
 - \frac{S_1^2}{2\pi} + \frac{S_1^2}{2\pi} \ln (4 p_f S_1) + \frac{\gamma}{\pi} S_1^2.
\label{8.10}
\end{equation}
Minimizing with respect to $S_0$ and $S_1$, we find  
\begin{eqnarray}
S_0\left[\ln(2 p_f)+ \gamma\right]- \gamma = 0,
\nonumber \\
S_1\left[ 2 \gamma + \ln (4 p_f S_1)\right]=0.
\label{8.11}
\end{eqnarray}
The first equation has the unique solution
\begin{equation}
S_0 = \frac{\gamma}{\gamma + \ln (2p_f)}.
\label{8.12}
\end{equation}
The 2nd equation has two solutions: $S_1=0$, corresponding to unbroken translational invariance,
and
\begin{equation}
S_1= \frac{1}{4 p_f}{\rm e}^{-2\gamma}
\label{8.13}
\end{equation}
for the soliton crystal. Comparing the energy densities of these two solutions,
\begin{equation}
E_{\rm g.s.}(S_1\neq 0) - E_{\rm g.s.}(S_1=0) = - \frac{1}{64\pi p_f^2} {\rm e}^{-4 \gamma},
\label{8.14}
\end{equation}
we learn that the crystal is favored, but the energy difference decreases rapidly with increasing $\gamma$.
We have shown this simple calculation in detail since it proves non-restoration of translational invariance at high
density for arbitrary bare fermion mass. 

\section{More about the double limit $T=0, \gamma=0$}

The simplest problem beyond individual baryons and the first one which could be solved analytically \cite{L30a}
is the ground state of baryonic matter in the chiral limit. 
The ansatz for the reduced HF potential has the form (\ref{7.1a}), whereas the scale factor $A$ is 
given by equation (\ref{8.2}). Thus only a single variational parameter $\kappa$ is left.
As expected, all analytical formulas of the previous section simplify tremendously in the chiral limit.
Thus for instance, the renormalized ground state energy density can now be expressed entirely in terms of complete 
elliptic integrals
\begin{equation}
\fl
E_{\rm g.s.} =   \frac{p_f^2 \mathbf{K}}{\pi^3}\left( 4 \mathbf{E} + (\kappa^2-2)\mathbf{K}\right)
+   \frac{2 p_f^2 \mathbf{K}}{\pi^3} \left(2\mathbf{E}
+(\kappa^2-2)\mathbf{K}\right) \ln \left(
\frac{\pi}{2 p_f  \kappa \mathbf{K}}\right) . 
\label{9.1}
\end{equation}
Minimizing with respect to $\kappa$  
yields the transcendental equation 
\begin{equation}
\kappa = \frac{a}{\ell} = \frac{\pi}{2 p_f \mathbf{K}},
\label{9.2}
\end{equation}
and self-consistency can be ascertained rather easily in this case.
Eliminating 
 $p_f$ from $E_{\rm g.s.}$ with the help of relation (\ref{9.2}), we arrive at the following 
parametric representation of the ground state energy as a function of density in
terms of the parameter $\kappa$,
\begin{eqnarray}
E_{\rm g.s.} &=& \frac{1}{4\pi} + \frac{1}{\pi \kappa^2} \left( \frac{\mathbf{E}}{\mathbf{K}}-\frac{1}{2}\right)\, ,
\label{9.3}
 \\
\frac{p_f}{\pi} &=& \frac{1}{2\kappa \mathbf{K}}\, .
\label{9.4}
\end{eqnarray}
In the low- or high-density limits, 
it becomes possible to systematically resolve the
transcendental equation (\ref{9.4}),
\begin{eqnarray}
\kappa  &  \begin{array}[t]{c} \approx \\
\raisebox{1ex} {\mbox{${\scriptstyle p_f \to \, 0}$}}
\end{array}  & 1 - 8 {\rm e}^{-\pi/p_f} +\frac{32(\pi + p_f)}{p_f}{\rm e}^{-2\pi/p_f}, 
\nonumber \\
\kappa  &  \begin{array}[t]{c} \approx \\
\raisebox{1ex} {\mbox{${\scriptstyle p_f \to \, \infty}$}}
\end{array} &
\frac{1}{p_f} - \frac{1}{4 p_f^3} + \frac{3}{64 p_f^5}. 
\label{9.8}
\end{eqnarray}
For the energy as a function of density, one then finds 
\begin{eqnarray}
E_{\rm g.s.}&  \begin{array}[t]{c} \approx \\
\raisebox{1ex} {\mbox{${\scriptstyle p_f \to \, 0}$}}
\end{array}  &  -\frac{1}{4\pi} + \frac{2 p_f}{\pi^2} + \frac{8 p_f}{\pi^2} {\rm e}^{-\pi/p_f}, 
\nonumber \\
E_{\rm g.s.}  &  \begin{array}[t]{c} \approx \\
\raisebox{1ex} {\mbox{${\scriptstyle p_f \to \, \infty}$}}
\end{array} & \frac{p_f^2}{2\pi} - \frac{1}{2^6 \pi p_f^2} + \frac{3}{2^{14}\pi p_f^6}. 
\label{9.9}
\end{eqnarray}
In the low-density limit, the three terms correspond to the vacuum energy density, the 
contribution from the baryon mass ($\sim \rho M_B$ with $M_B=2/{\pi}$) and a term reflecting
the repulsive baryon-baryon interaction. At high densities, we can identify the free massless
Fermi gas piece, the leading perturbative correction \cite{L30} and the next term
coming from higher order effects, suggesting fast convergence. 

We end this section with a few remarks about the baryon density. In the present case,
the exact density is $x$-dependent and given by
\begin{equation}
\rho(x) = \frac{1}{2\kappa\mathbf{K}} - \frac{\mathbf{K}'}{2\pi\kappa}\left(\tilde{S}^2(x/\kappa)-\langle \tilde{S}^2 \rangle \right) \, .
\label{9.5}
\end{equation}
It has the following high and low-density limits: At low density ($\kappa \to 1$) we recover the result
for a single baryon
\begin{equation}
\rho(x,\kappa \to 1) \approx \frac{1}{4\cosh^2x}+\frac{1}{4 \cosh^2(x/\kappa+\mathbf{K})} \, .
\label{9.6}
\end{equation}
At high density ($\kappa \to 0$) the total baryon density approaches a constant,
\begin{equation}
\rho(x,\kappa \to 0) \approx \frac{1}{2\kappa\mathbf{K}} ,
\label{9.7}
\end{equation}
unlike the scalar potential which keeps oscillating around 0 with wave number $2p_f$.

\section{A note on baryons}

Single baryons are contained in the low temperature, low density limit of the general formalism, at least those with
completely filled or empty valence state. They have also been studied as such,
first in the chiral limit \cite{L8} and more recently in the massive GN model \cite{L18,L19}.
One finds that the scalar potential has the same shape as the $m_0=0$ baryon with partial filling of the valence level,
\begin{equation}
S(x) =   1 + y \left( \tanh \xi_-  - \tanh \xi_+ \right) 
\label{10.1}
\end{equation}
with
\begin{equation}
\xi_{\pm} = y x \pm \frac{1}{2} {\rm artanh}\, y ,
\label{10.2}
\end{equation}
where $m=1$ and $y\in [0,1]$ is the only variational parameter.
It depends on the occupation of the pair of valence levels and $\gamma$.
Evaluating the ground state energy $M_B$ and varying with respect to $y$,
\begin{equation}
\frac{\partial M_B}{\partial y }  = 0,
\label{10.3}
\end{equation}
one obtains
\begin{equation}
\frac{\nu}{2} = \frac{\theta}{\pi} + \frac{\gamma}{\pi} \tan \theta  
\label{10.4}
\end{equation}
where we have introduced the angle $\theta$ via $y=\sin \theta$ ($0 \leq \theta \leq \pi/2$). 
Here $\nu$ is the number of valence particles minus the number of holes in the negative energy valence state.
The baryon mass at the minimum becomes
\begin{equation}
\frac{M_B}{N} = \frac{2}{\pi} \left[ \sin \theta + \gamma \,{\rm artanh}  (\sin \theta) \right].
\label{10.5}
\end{equation}
The phase boundary in the $T=0$ plane of figure 5 agrees perfectly with the function $M_B(\gamma)$ evaluated from
equations (\ref{10.4},\ref{10.5}) for $\nu=1$. This was not the case in the old phase diagram of figure 2,
confirming once again that the revised results are now internally consistent.

\section{Relation to condensed matter physics}

From the point of view of relativistic QFT, the GN model  is viewed as a soluble model
for strong interaction physics, exhibiting phenomena like asymptotic freedom, chiral symmetry breaking, 
dynamical mass generation, meson and baryon bound states.  
On the other hand, it can model quasi-one-dimensional condensed
matter systems  in the vicinity of a half-filled band. In this context, it has enjoyed success in
explaining real experimental data, something unthinkable in particle physics due to its toy model character.
The two most striking examples which we are aware of are conducting polymers on the one hand
and (quasi-one-dimensional) inhomogeneous superconductors on the other hand. The basic physics is the same
in both cases and closely
related to the Peierls effect \cite{L31} of a one-dimensional electron-phonon system --- dynamical formation of a gap at 
the Fermi surface resulting in a crystal structure. Details are
different though, in particular the identification of the chemical potential is more subtle in the second case.
Since the relationship between the GN model and condensed matter systems has played an important role
for establishing the phase diagram of the GN model during the last few years, let us discuss it in 
somewhat more detail.
\vskip 0.2cm

\noindent {\em 1. The GN model and polyacetylene}

Conducting polymers have had a tremendous success story, culminating in the 2000 Nobel Prize in chemistry
for physicist A. J. Heeger and the chemists A. G. MacDiarmid and H. Shirakawa (for a review, see \cite{L32}). 
A prominent example is {\em trans}-polyacetylene (PA). This polymer (CH)$_x$ possesses two ``dimerized",
degenerate ground states with alternating short and long bonds (figure 12, left). Owing to a number of simplifying
assumptions, its continuum description is mathematically
equivalent to the symmetric ($m_0=0$) GN model, as was realized shortly after the seminal work of Su, Schrieffer
and Heeger on the discrete model \cite{L33,L9,L35,L36}.
\begin{figure}[ht]
\begin{center}
\epsfig{file=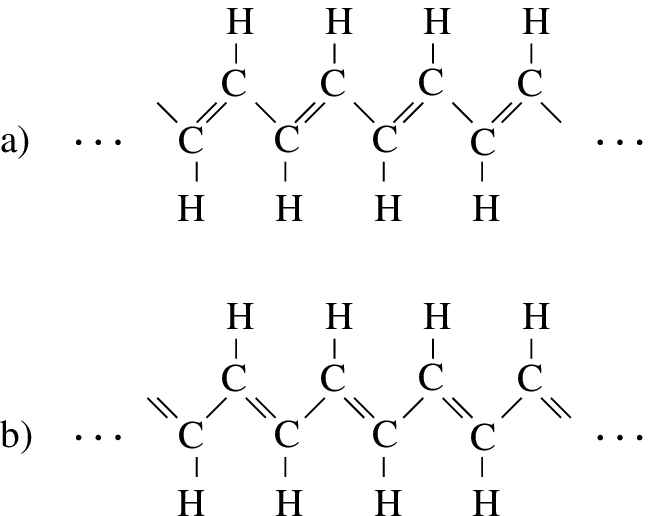,width=6cm}\hskip 1.0cm \epsfig{file=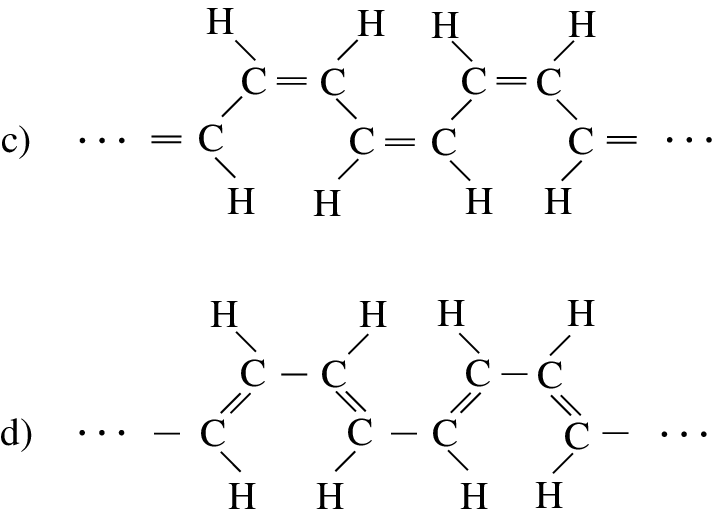,width=6cm}
\caption{Left: Two degenerate, dimerized ground states a) and b) of {\em trans}-PA, described theoretically in terms
of the massless GN model. Right: Two inequivalent configurations of {\em cis}-PA, leading to a description in terms of 
the massive GN model. Configuration c) has lower energy than d).}
\end{center}
\end{figure}
Dimerization plays the role of (discrete) chiral symmetry breaking in relativistic QFT.
Solitons appear as kink-like defects where a transition between the two 
degenerate ground states happens, polarons are kink-antikink bound states. These excitations  
are important for understanding the electrical conductivity properties of doped PA (doping changes the number
of electrons and is analogous to changing baryon number in the GN model).
The solitons in particular have attracted considerable attention in the context of fermion number
fractionization of which they are a prime example \cite{L37,L38,L39}.           
Solitons and polarons have originally been analyzed theoretically
in close analogy with kink and kink-antikink baryons already known on the field theory side \cite{L8}.
A good account of both the physics and the early history can be found in reference \cite{L40}, where also the limitations 
of the equivalence between the GN model and the theory of PA are critically examined. More recently, this theme has been
taken up again by others \cite{L41}.
Here, let us only mention a few facts which help to understand the curious equivalence of a non-relativistic
electron-phonon system with a relativistic QFT. Particle/antiparticle degrees of freedom correspond to particles 
and holes, the half filled band to the Dirac sea. The role of the UV cutoff is taken over by
the band width $W$ (incidentally of the order of 10 eV in PA). The two spin-components (which exist only in the condensed
matter case) are mapped onto 2 flavors (i.e., the $N=2$ GN model). In the widely used semi-classical
approach, there is no difference between the condensed matter treatment and the large $N$ limit used above.
The linear (``ultra-relativistic") dispersion relation for the massless fermions 
arises from a linearization at the Fermi surface, the Fermi velocity playing the role of the velocity of light.
The lattice distortion (or phonon) field in the adiabatic approximation corresponds to the auxiliary scalar field
$\sigma$ in the GN model, the gap parameter $\Delta(x)$ to the scalar potential $S(x)$ 
and the electron-phonon coupling $\lambda$ to the coupling constant $Ng^2$.
In the relativistic QFT case, one has to send the UV cutoff to infinity and the bare parameter 
$Ng^2$ to 0 in a specific way, dictated by the renormalizability of the model. In polymer physics,
$W$ and $\lambda$ are physical observables to be taken from experiment. Nevertheless, in practice
these differences do not matter for many questions. 

As mentioned above, the kink baryons are the solitons, the kink-antikink baryons the polarons, bipolarons and excitons of 
polymer physics. More specifically, for $N=2$
the valence level can only be empty, fully occupied or half occupied. In condensed matter physics, there
is then a relation between charge and spin. For the questions
discussed in the present paper, fully occupied or empty valence levels are most relevant --- these should be
identified with the 
bipolarons. In the massless GN model, the baryon with fully occupied valence level becomes a kink/antikink at
infinite separation; this is the reason why bipolarons are not discussed in the case of degenerate
polymers like {\em trans-}PA.

If one starts to think about the massive GN model, it is not too hard to identify its 
condensed matter analogue: Conducting polymers with non-degenerate ground states
like {\em cis}-PA (figure 12, right).
Their theoretical description was initiated by Brazovskii and Kirova \cite{L42} with the proposal
that the gap parameter has two contributions, a constant, ``external"  one arising from the basic structure of the 
polymer and an $x$-dependent, ``internal" one due to electron-phonon coupling,
\begin{equation}
\Delta (x) = \Delta_e + \Delta_i(x)  .
\label{11.1}
\end{equation} 
If we identify $\Delta(x)$ with the scalar mean field $S(x)$ and $\Delta_e$ with the bare mass $m_0$, we can immediately 
relate two problems from two different branches of physics. Baryons in the massive GN model then correspond to 
polarons and bipolarons in the polymer case \cite{L40,L18}. 

Now consider periodic solutions of the gap parameter for both degenerate and non-degenerate polymers and 
compare them with relativistic QFT. In the degenerate polymer case, polaron crystal
solutions to the continuum model (at $T=0$) have been found and discussed in \cite{L43,L44,L45}.
They are mathematically identical to the ground state of baryonic matter in the chirally symmetric
GN model, cf. section 9.
In some cases the order parameter looks different from ours at first sight. However, the two expressions can be
converted into each other by Landen's transformation for Jacobi functions in the form
\begin{equation}
\kappa \frac{{\rm sn}(\xi,\kappa) {\rm cn}(\xi,\kappa)}{{\rm dn}(\xi,\kappa)}
= \frac{1-\kappa'}{\kappa} {\rm sn} \left( (1+\kappa')\xi, \frac{1-\kappa'}{1+\kappa'} \right) .
\label{11.2}
\end{equation} 
In the 80's and 90's, a lot of work was devoted to non-degenerate conducting polymers with the result that exact
bipolaron lattice solutions were found by several groups. This suggests that the massive GN model
should exhibit a soliton crystal at finite baryon density as well. It encouraged us 
to look for such solutions of the massive GN model and to try the functional form of the self-consistent scalar
potential for non-degenerate polymers in the GN model. It has been derived by various methods such as inverse
spectral theory \cite{L46}, Poisson
summation of periodic sums of single polarons \cite{L47,L48,L49} or 
relation to the Toda lattice \cite{L22} --- a rich source of inspiration also for those working on relativistic QFT.
Reference \cite{L23} contains a particularly thorough discussion and corrects
misprints in some of the original papers. In section 3 we have in fact presented the functional form determined
in these works and used it as ansatz in a relativistic HF calculation.
The arguments given in favor of this ansatz are also borrowed from the condensed matter literature. 
On the other hand we did not find the full phase diagram at finite $T$ presented in section 5 in the polymer literature.
\vskip 0.2cm

\noindent {\em 2. The GN model and inhomogeneous superconductors}

Superconductivity is driven by fermion-fermion pairing (Cooper pairs), whereas the GN model features 
fermion-antifermion pairing (chiral symmetry breaking). We first have to understand how 
these two distinct physical phenomena are related to each other.
Along the lines described in reference \cite{L50}, one can actually map the 
GN Lagrangian onto a ``dual" Lagrangian which has fermion-fermion pairing by means of a canonical 
transformation. 
In the relativistic case, this is possible due to a 2-dimensional remnant of the Pauli-G\"ursey symmetry of massless
fermions \cite{L51,L52} and explains why two seemingly different large $N$ field theory models
give identical results \cite{L53}.
All one has to do is redefine particles into antiparticles for left-handed fermions only. If one works at
non-zero chemical potential, a baryonic chemical potential $\mu$ in the GN model corresponds to
an ``axial" chemical potential $\mu_5$ in the dual BCS-type model. 
The phase diagram which we have discussed in section 5 (for $m_0=0$) is equivalent to the phase diagram
of a theory with Lagrangian
\begin{equation}
{\cal L} = \sum_{n=1}^N \bar{\psi}^{(n)} {\rm i} \partial \!\!\!/ \psi^{(n)} + \frac{g^2}{2} \left[ \,\sum_{n=1}^N\left( \psi^{(n)\dagger}_R 
\psi^{(n)\dagger}_L + \psi^{(n)}_L \psi^{(n)}_R \right)\right]^2  ,
\label{11.3}
\end{equation}
provided we reinterpret the chemical potential $\mu=\mu_R+\mu_L$ as 
axial chemical potential $\mu_5 = \mu_R-\mu_L$. The kink-antikink phase of the GN model
can then be identified with
the Larkin-Ovchinnikov-Fulde-Ferrel (LOFF) phase \cite{L54,L55} of the dual model. 
Such inhomogeneous superconductors have recently attracted considerable attention in the context of QCD (for a 
review article, see reference \cite{L56}). 

Let us now turn to non-relativistic condensed matter physics and demonstrate that the GN model can also be 
related to quasi-one-dimensional superconductors in nature.
In 1981, Mertsching and Fischbeck addressed the quasi-one-dimensional Peierls-Fr\"ohlich model
with a nearly half-filled band,
an electron-phonon system \cite{L57}. This is the same basic model as the continuum model for
degenerate polymers \cite{L9}, but here the phase diagram at finite temperature 
was considered, notably the transition between commensurate-incommensurate charge density waves.
Comparing with our results at $m_0=0$, we find  
a mathematical one-to-one correspondence between this system and the 
GN model now extended to finite temperature. The authors of reference \cite{L57} have also found the
analytic solution to the mean field equation, guided by the Landau expansion around the triple point
(which is called Leung point \cite{L58} in this context).  

In a subsequent paper, Machida and Nakanishi \cite{L59} used the phase diagram of reference \cite{L57} in a
different physics context: They studied the interplay of superconductivity and ferromagnetism
in ErRh$_4$B$_4$ (Erbium-Rhodium-Boride). They managed to reduce this problem 
mathematically to the Peierls-Fr\"ohlich model. For real order parameter, their results are again
fully equivalent to ours for the GN model, except that now 
one has to use another dictionary: The Dirac equation corresponds to the Bogoliubov-deGennes (BdG) equation,
flavor ($N=2$) to spin (which exists in a quasi-one-dimensional world), chemical
potential to magnetic field, baryon density to spin polarization. Our three phases (massive,
crystal and massless) correspond to their BCS, ``sn" and normal phases, respectively. 
Not only the phase boundaries, but all observables can be identified if one keeps in mind the above mentioned dictionary.
Let us try to understand in more detail the reasons behind this remarkable correspondence. The BdG Hamiltonian
involves four fermion fields,
namely right ($\psi_s$) and left ($\phi_s$) moving electrons with spin up ($s=+$) and spin down ($s=-$).
This Hamiltonian can be mapped onto the Peierls-Fr\"ohlich Hamiltonian by the canonical transformation
\begin{equation}
\left( \begin{array}{c} \psi_+ \\ \psi_- \\ \phi_+ \\ \phi_- \end{array} \right) \longrightarrow
\left( \begin{array}{c} \psi_+ \\ \psi_-^{\dagger} \\ \phi_- \\ \phi_+^{\dagger} \end{array} \right),
\label{11.4} 
\end{equation}
i.e., particle hole conjugation for spin-down fields followed by a spin-flip of left-moving fields. Under this transformation,
spin density goes over into ordinary fermion density, and one can understand all other relationships as well.

Finally, we would like to mention the more recent work of Buzdin and Kachkachi \cite{L60}. They derive the 
Ginzburg-Landau 
theory for nonuniform LOFF superconductors near the tricritical point in the ($T,H$)-phase diagram in one, two
and three dimensions. If we take their result for one dimension and specialize it to a real order parameter, we 
find perfect agreement between our equation (\ref{6.5}) and their equation (3) in appropriate units. 
Once again we have to identify their magnetic field ${\cal H}_0$ with our chemical potential $\mu$ for
the reasons discussed above. 

\section{Concluding remarks}

At the time of writing our previous review article on the thermodynamics of two-dimensional quantum field theories
\cite{L5}, some progress had been made on fermionic models with continuous chiral symmetry
like the 't~Hooft or the NJL$_2$ model.
The fact that the existence of baryons implied a crystal structure of baryonic matter had been understood and 
the physical interpretation in terms of a gap at the Fermi surface --- the ``rediscovery of the Peierls effect in
relativistic QFT" --- had been clearly stated. Towards the end of
this article we identified areas where future work was needed, mentioning in particular the GN model with discrete
chiral symmetry. We hope to have shown in the present work that this problem has been solved in the meantime,
including the generalization to the massive model with explicit symmetry breaking. As a result, the
situation is now reversed: Today we know a lot more about the thermodynamics of the discrete chiral GN model
than about the NJL$_2$ model, particularly in the presence of a bare mass term.
For a long time the phase diagram of the NJL$_2$ model was thought to be identical to that of the GN model, but
this can now be ruled out due to their different baryon structure. 
It will be interesting to see whether one can make further progress on the phase diagram of the massive
NJL$_2$ model where topology is expected to play a crucial role, analogous to the Skyrme model 
and Skyrme crystal in 3+1 dimensions. 

In reference \cite{L5} we also pointed out that this type of soluble QFT models still had the potential 
to surprise us after so many years of studies. If anything, this impression has only been reinforced since then. 
The simple Lagrangian (\ref{1.1}) has generated a much richer phase diagram than previously thought,
together with some beautiful mathematics. Even more surprising for us was perhaps the 
discovery that a relativistic QFT ``toy model" has such a close relationship to various quasi-one-dimensional
condensed matter systems. We have tried to cover these less familiar aspects of the GN model 
in the present work as well. They represent a good example for a fruitful exchange between condensed matter
and particle physics. Originally, condensed matter physics, in particular polymer physics, could profit
of the particle physicists know-how on soliton-like baryons in the GN model to understand for instance electrical
conductivity properties. Subsequently the polymer continuum models were vigorously developed, notably in the
direction of crystal solutions where the expertise has traditionally been residing in condensed matter physics.
During our work, we were able to take advantage of the progress
in polymer physics to settle some unresolved issues in QFT, while at the same time generalizing the 
bipolaron lattice to finite temperature. 

This whole process was not as straightforward
as it may sound. It was not that easy for us to find the relevant information in the unfamiliar
literature where it was often hidden under a lot of material of less interest to 
us. Thus  we could not avoid doing many things ``the hard way". For example the calculation of the phase diagram
in the chiral limit, first numerically and then analytically, was performed by us independently of 
already existing pertinent results in the ``parallel world" of condensed matter physics. In any case,
our study of the massive model 
has profited immensely from this interplay, and it was better to realize and exploit the relationship between 
these two different branches of physics late than never.

\vskip 0.5cm
\noindent
The author wishes to thank the guest editors of this special issue of J.~Phys.~A, in particular Gerald Dunne, for the
invitation to write this contribution. A most pleasant and productive collaboration with Oliver Schnetz and Konrad
Urlichs on the work reported here is gratefully acknowledged. I also thank both of my collaborators for a critical
reading of the manuscript.

\end{document}